# The construction of a universal quantum gate set for the $SU(2)_k$ (k=5,6,7) anyon models via genetic optimized algorithm

Jiangwei Long [1], Yizhi Li [2], Jianxin Zhong [3, 4] and Lijun Meng [1, 4, †]

[1] School of Physics and Optoelectronics, Xiangtan University, Xiangtan 411105, Hunan, People's Republic of China
[2] School of Physics and Electronic Science, Hunan Institute of Science and Technology, Yueyang 414006, People's Republic of China
[3] Center for Quantum Science and Technology, Department of Physics, Shanghai University, Shanghai 200444, People's Republic of China
[4] Hunan Key Laboratory for Micro-Nano Energy Materials and Devices, Hunan, People's Republic of China

**Abstract** We study systematically numerical method into constructing a universal quantum gate set for topological quantum computation (TQC) using $SU(2)_k$ anyon models. The *F*-matrices and *R*-symbol were computed through the *q*-deformed representation theory of SU(2), enabling precise determination of elementary braiding matrices (EBMs) for $SU(2)_k$ anyon systems. Quantum gates were derived from these EBMs. One-qubit gates were synthesized using a genetic algorithm-enhanced Solovay-Kitaev algorithm (GA-enhanced SKA), while two-qubit gates were constructed through brute-force search or GA optimization to approximate local equivalence classes [CNOT]. Implementing this framework for $SU(2)_5$, $SU(2)_6$, and $SU(2)_7$ models successfully generated the canonical universal gate set {*H*-gate, *T*-gate, CNOT-gate}. These numerical results provide conclusive verification of the universal quantum computation capabilities inherent in $SU(2)_k$ anyon models. Furthermore, we get exact implementations of the local equivalence class [SWAP] using nine EBMs in each model.

## 1 Introduction

The TQC fundamentally relies on the braiding statistics of non-Abelian anyons. The foundational proposal for harnessing non-Abelian anyons in TQC was first established by A.Yu. Kitaev [1]. The anyon was first conceptualized in 2D quantum systems by Myrheim and Leinaas [2], non-Abelian anyons are specifically characterized by their multidimensional fusion channels (Non-Abelian anyons have multiple possible resulting states during their fusion.) and non-commutative braiding properties, contrasting sharply with Abelian anyons that exhibit single-dimensional fusion outcomes and commutative statistics [3-7]. The principal advantage of TQC over non-TQC lies in its intrinsic fault tolerance - the topological nature of information encoding provides inherent protection against local noise perturbations [8,9]. Experimental realization of TQC necessitates the physical manifestation of non-Abelian anyonic excitations, consequently driving sustained research efforts in condensed matter systems ranging from fractional quantum Hall states to topological superconductors[10-20].

† Corresponding author. E-mail: ljmeng@xtu.edu.cn

In $SU(2)_k$ models, the k=2 case corresponds to the Ising anyon model. While Ising anyons cannot achieve universal quantum computation through braiding operations alone due to the impossibility of implementing the *T*-gate via braiding [21], their physical realization as Majorana fermions remains the most experimentally accessible candidate for non-Abelian anyons. Recent studies have indicated that, the introduction of a so-called "neglecton" into the conventional Ising anyon model enables universal quantum computation solely through braiding operations within the modified Ising anyon model based on non-semisimple topological quantum field theory [22]. The k=3 case represents the Fibonacci anyon model – the simplest known non-Abelian system enabling universal quantum computation purely through braiding operations [23]. Extensive theoretical work has demonstrated the capability of Fibonacci anyons to construct fundamental quantum gates spanning one-qubit [24,25], two-qubit [26-28], three-qubit [29], and generalized *N*-qubit operations [30]. At k=4, the metaplectic anyon model requires supplementary measurement and fusion protocols to attain computational universality, as braiding operations alone prove insufficient for this implementation [31,32]. Recently, Sergey Mironov and Andrey Morozov proposed using the cabling method in knot theory to regard the anyon pairs of two qubits as "cables", braid them to realize low-leakage two-qubit entangled gates in TQC [33]. In previous work, we constucted a universal set of quantum gates for the metaplectic anyon model by braiding and fusion operations using unconventional encoding strategies [34]. Theoretical analyses confirm that $SU(2)_k$ models with k>3 (k≠4) achieve dense coverage of the SU(2) group for universal quantum computation [35]. However, despite numerical verification of one- and two-qubit gate implementations in the Fibonacci model (k=3) [25,28], no numerical evidence currently supports the existence of complete universal gate sets in $SU(2)_k$ models with k>4.Using the *q*-deformed representation theory of SU(2) [36], which introduced the parameter *q* to generalize the classical representation theory of SU(2), we derived the *F*-matrix (Fusion matrix, which corresponds to the basis transformation between different fusion states.) and *R*-symbol (Rotation symbol, which corresponds to the phase acquired from once braiding of two anyons.) for $SU(2)_k$ anyon models. From these symbols, the EBMs (Some unitary matrices that represents the exchange of two anyons.) for both one-qubit and two-qubit configurations were analytically determined. Subsequently, we constructed a universal quantum gate set {*H*-gate, *T*-gate, CNOT-gate} [37] through strategic implementations of these EBMs. This provides the first numerical demonstration that $SU(2)_k$ models with k > 4 can indeed achieve universal quantum computation. For concrete demonstration, we obtained the EBMs for $SU(2)_5$, $SU(2)_6$, and $SU(2)_7$ systems explicitly. One-qubit gates {*H*-gate, *T*-gate} were synthesized using our GA-enhanced SKA, while the local equivalence class [CNOT] (Local equivalence classes refer to the equivalence relation wherein two quantum two-qubit gates are interconvertible via one-qubit operations.) was approximated through exhaustive search or GA optimization. Numerical simulations reveal high-fidelity implementations of the {*H*-gate, *T*-gate, CNOT-gate} through these EBMs, with computational precision comparable to that achieved in Fibonacci anyons.

Section 2 details the encoding architectures for one- and two-qubit systems using $SU(2)_5$, $SU(2)_6$, and $SU(2)_7$ non-Abelian anyons, along with GA-enhanced SKA methodology for quantum gate compilation. Section 3 presents numerical implementations of the {*H*-gate, *T*-gate, CNOT-gate} constructed through our framework, accompanied by fidelity metrics and computational benchmarks. Section 4 provides conclusion. Appendix A contains the mathematical framework of *q*-deformed SU(2) representation theory employed for deriving *F*-symbols and *R*-symbols. Appendix B provides explicit *F*-matrix and *R*-symbol solutions used to determine EBMs in our implementations. Appendix C outlines the generalized computational workflow for obtaining EBMs in $SU(2)_k$ anyon systems. Appendix D determines which topological spin particles we select to construct topological quantum gates for $SU(2)_{5,6,7}$ anyon models. Appendix E presents the anyon world-lines corresponding to the braidwords provided in this paper.

## 2 Models and methods

The implementation of qubits through non-Abelian anyons necessitates a fusion protocol governed by twofold degeneracy, according to the *k*-level theory [38]:

$$s_1 \otimes s_2 = |s_1 - s_2| \oplus |s_1 - s_2| + 1 \oplus ...... \oplus \min(s_1 + s_2, k - s_1 - s_2) \tag{1}$$

where $\otimes$ represents the fusion operation, and $\oplus$ denotes the combination of possible fusion outcomes. The fusion of anyons with topological spins $s_1$ and $s_2$ produces resultant anyons whose topological spins start from $|s_1-s_2|$, increment sequentially by 1, and terminate at the minimum value between $s_1+s_2$ and $k-s_1-s_2$.

The composition of a qubit conventionally employs non-Abelian anyons with topological spin-1/2, as two such particles inherently satisfy the required fusion rule $\frac{1}{2} \otimes \frac{1}{2} = 0 \oplus 1$. Appendix D discusses the reasons for our choice of spin-1/2 particles to construct topological quantum gates rather than other higher-spin particles. This equivalence implies identical computational capabilities for qubit realizations across these distinct topological spin configurations. One-qubit encoding permits two equivalent schemes: 3-anyons or 4-anyons configurations [25]. We adopt the 3-anyons architecture for its dimensional advantage: compared to the 4-anyons case, it reduces the number of fusion channels for a two-qubit system, thereby lowering the dimensionality of the two-qubit EBMs and substantially facilitating the analytical determination of their elements.

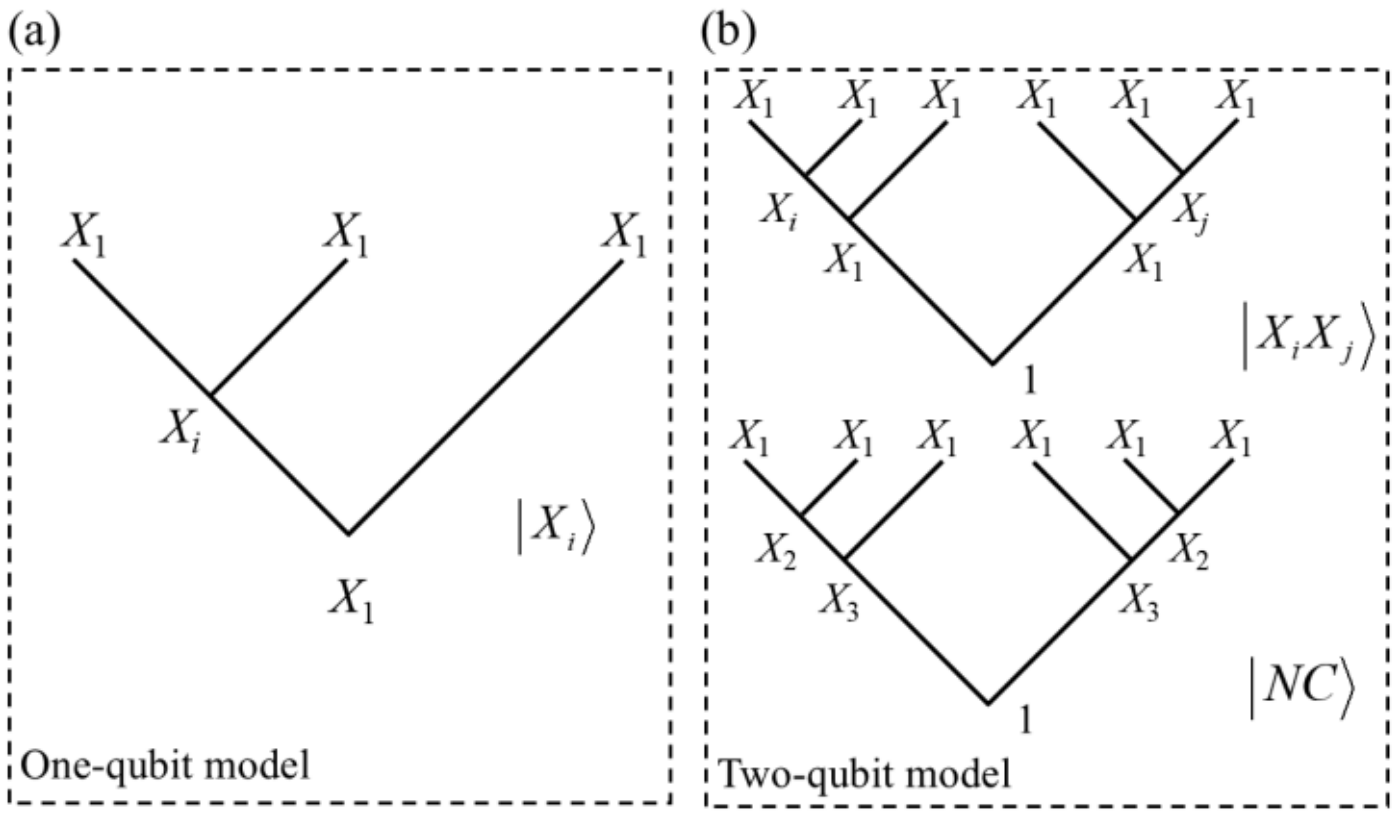

**Fig. 1**: (a) Schematic diagram of a one-qubit encoding scheme utilizing three topological spin-1/2 anyons. (b) Two-qubit encoding architecture employing six topological spin-1/2 anyons, with the computational states (upper configuration) and non-computational states (lower configuration).

We employ doubled topological spin values to label individual anyons. As illustrated in Fig. 1(a), the one-qubit encoding scheme utilizes three topological spin-1/2 anyons (denoted $X_1$ with subscript double spin 1). The fusion protocol proceeds sequentially: initial fusion of two $X_1$ anyons yields either vacuum 1 or $X_2$(Corresponding to a topological spin-1 anyon), followed by subsequent fusion with the third $X_1$ to finalize the $X_1$ outcome. The intermediate fusion state $|1\rangle/|X_2\rangle$ corresponds to the logical qubit state $|0\rangle/|1\rangle$. Similarly, as showed in Fig. 1(b), the two-qubit architecture employs six topological spin-1/2 anyons, where the intermediate fusion state $|11\rangle/|1X_2\rangle/|X_21\rangle/|X_2X_2\rangle$ maps to the logical state $|00\rangle/|01\rangle/|10\rangle/|11\rangle$. However, this process of fusion introduces an additional non-computational state. As shown in lower Fig. 1(b), the $|NC\rangle$ state arises from an additional intermediate state that differs from the four intermediate states corresponding to the computational basis. Specifically, for six initial particles with topological spin-1/2, the first two and the last two topological spin-1/2 particles can first fuse into a topological spin-1 particle. These then fuse individually with the third topological spin-1/2 particle to form two topological spin-3/2 particles. Finally, the fusion of these two topological spin-3/2 particles can still yield the vacuum state 1. Consequently, the EBMs for two-qubit operations manifest as 5-dimensional matrices, with the computational subspace embedded within this extended space.

The EBMs for $SU(2)_5$, $SU(2)_6$, and $SU(2)_7$ models were derived following this workflow:

① Numerical evaluation of $F$- matrices and $R$-matrices using the $q$-deformed SU(2) representation theory framework (formulae provided in Appendix A).

② Systematic construction of EBMs by implementing braiding operations through sequential F-moves and R-moves, with operator projected onto each computational basis.

Explicit numerical values of the $F$- and $R$- matrices employed in EBMs derivations are cataloged in Appendix B. A stepwise protocol for EBMs determination is presented in Appendix C.

Within the computational basis $\{|1\rangle, |X_2\rangle\}$, the one-qubit EBMs take the form:

$$\mathrm{SU}(2)_5:$$

$$\sigma_1^{(3)} = \begin{bmatrix} -0.78183148+0.62348980i & 0 \\ 0 & 0.97492791+0.22252093i \end{bmatrix}$$

$$\sigma_2^{(3)} = \begin{bmatrix} 0.43388374+0.34601074i & 0.81102135-0.18511033i \\ 0.81102135-0.18511033i & -0.24078731+0.5i \end{bmatrix}$$

$$\mathrm{SU}(2)_6:$$

$$\sigma_1^{(3)} = \begin{bmatrix} -0.83146961+0.55557023i & 0 \\ 0 & 0.98078528+0.19509032i \end{bmatrix}$$

$$\sigma_2^{(3)} = \begin{bmatrix} 0.44998811+0.30067244i & 0.82473883-0.16405075i \\ 0.82473883-0.16405075i & -0.30067244+0.44998811i \end{bmatrix}$$

$$\mathrm{SU}(2)_7:$$

$$\sigma_1^{(3)} = \begin{bmatrix} -0.86602540+0.5i & 0 \\ 0 & 0.98480775+0.17364818i \end{bmatrix}$$

$$\sigma_2^{(3)} = \begin{bmatrix} 0.46080249+0.26604444i & 0.83382540-0.14702592i \\ 0.83382540-0.14702592i & -0.34202014+0.40760373i \end{bmatrix}$$

Within the non- computational and computational basis $\{|NC\rangle, |11\rangle, |1X_2\rangle, |X_21\rangle, |X_2X_2\rangle\}$, the two-qubit EBMs take the form:

$$\sigma_1^{(6)} = R_2^{11} \oplus \left(\sigma_1^{(3)} \otimes I_2\right), \sigma_2^{(6)} = R_2^{11} \oplus \left(\sigma_2^{(3)} \otimes I_2\right),$$
$$\sigma_4^{(6)} = R_2^{11} \oplus \left(I_2 \otimes \sigma_2^{(3)}\right), \sigma_5^{(6)} = R_2^{11} \oplus \left(I_2 \otimes \sigma_1^{(3)}\right),$$

$$\mathrm{SU}(2)_5:$$

$$\sigma_3^{(6)} = \begin{bmatrix} 0.44504187i & 0 & 0 & 0 & 0.87305746-0.19926967i \\ 0 & -0.78183148+0.62348980i & 0 & 0 & 0 \\ 0 & 0 & 0.97492791+0.22252093i & 0 & 0 \\ 0 & 0 & 0 & 0.97492791+0.22252093i & 0 \\ 0.87305746-0.19926967i & 0 & 0 & 0 & 0.19309643+0.40096887i \end{bmatrix}$$

$$\mathrm{SU}(2)_6:$$

$$\sigma_3^{(6)} = \begin{bmatrix} -0.08080906+0.40625456i & 0 & 0 & 0 & 0.89269087-0.17756725i \\ 0 & -0.83146961+0.55557023i & 0 & 0 & 0 \\ 0 & 0 & 0.98078528+0.19509032i & 0 & 0 \\ 0 & 0 & 0 & 0.98078528+0.19509032i & 0 \\ 0.89269087-0.17756725i & 0 & 0 & 0 & 0.23012473+0.34440599i \end{bmatrix}$$

$\mathrm{SU}(2)_7$ :

$$\sigma_3^{(6)} = \begin{bmatrix} -0.13507430+0.37111360i & 0 & 0 & 0 & 0.90475357-0.15953247i \\ 0 & -0.86602540+0.5i & 0 & 0 & 0 \\ 0 & 0 & 0.98480775+0.17364818i & 0 & 0 \\ 0 & 0 & 0 & 0.98480775+0.17364818i & 0 \\ 0.90475357-0.15953247i & 0 & 0 & 0 & 0.25385665+0.30253458i \end{bmatrix}$$

The operator $\sigma_i^{(n)}$ denotes the braiding of the $i$-th and ($i$+1)-th anyons, where the superscript ($n$=3, 6) specifies the encoding architecture: $\sigma_i^{(3)}$ acts on 3-anyon one-qubit systems, while $\sigma_i^{(6)}$ operates on 6-anyon two-qubit configurations. This superscript notation explicitly distinguishes between one- and two-qubit EBMs. All EBMs rigorously satisfy the Artin braid group relations [39]:

$$\begin{aligned} &\sigma_i\sigma_j = \sigma_j\sigma_i \quad \text{for } |i-j| \geq 2, \\ &\sigma_i\sigma_{i+1}\sigma_i = \sigma_{i+1}\sigma_i\sigma_{i+1}. \end{aligned} \tag{2}$$

This implies that the braiding processes inherently exhibit topological protection.

We now address the construction of a universal quantum gate set {$H$-gate, $T$-gate, CNOT-gate} using the numerically derived EBMs.

For Fibonacci anyon-based one-qubit gate compilation – a paradigmatic quantum compiling challenge – braidword formed by EBM sequences approximate target unitary gates in an exponentially large space. Established methodologies include the SKA [40], hash function techniques [41], GA [42], algebraic techniques [43], reinforcement learning [44], and Monte Carlo-enhanced SKA [45]. The quantum compiling problem for $\mathrm{SU}(2)_{5,6,7}$ anyon-based one-qubit gates shares analogous structure with the Fibonacci case. To solve this, we employ our previous developed GA -enhanced SKA [46]. Prior implementations on Fibonacci anyons demonstrated superior performance of GA -enhanced SKA over Monte Carlo-enhanced SKA. For completeness, we outline the method below:

The approximation error between generated braidword and target gates is quantified using the global-phase-invariant distance metric [47]:

$$d\left(U_0, U\right) = \sqrt{1 - \frac{\left|Tr\left(U_0 U^{\dagger}\right)\right|}{2}}, \tag{3}$$

where $U_0$ denote the unitary matrix representation of the braidword, $U$ the target one-qubit gate, and $Tr$ the trace of matrix. The metric asymptotically approaches 0 as $U_0$ converges to $U$, up to a global phase. This phase-invariant formulation explicitly disregards global phase differences – a physically inconsequential factor in quantum computation.

The Solovay-Kitaev algorithm provides an efficient method to approximate any desired quantum gate (The desired quantum gate are the standard *H*- and *T*-gates in this context) with high precision using a finite set of basic gates (The finite set of basic gates are the EBMs corresponding to the $SU(2)_{5,6,7}$ anyon model in this context). This algorithm ensures that any quantum operation can be synthesized efficiently and with minimal overhead.

The canonical SKA employs the following pseudocode framework:

```
function Solovay-Kitaev(Gate U, depth n)
if (n == 0)
   Return Basic Approximation to U
else
   Set U_{n-1} = Solovay-Kitaev(U, n-1)
   Set V, W = GC-Decompose(U U†_{n-1})
   Set V_{n-1} = Solovay-Kitaev(V, n-1)
   Set W_{n-1} = Solovay-Kitaev(W, n-1)
   Return U_n = V_{n-1} W_{n-1} V†_{n-1} W†_{n-1} U_{n-1}
```

The algorithm recursively generates approximations $U_n$ to the target gate $U$, progressively minimizing the operator distance at the expense of a fivefold increase in braid length (number of EBMs) and threefold temporal overhead per recursion level. The standard SKA implementation proceeds as follow:

(i) 0-level approximation $U_0$: obtained via exhaustive search directly.

(ii) 1-level approximation $U_1$: setting $\Delta = UU_0^\dagger$ and performing group commutator decomposition (GC-decomposition) $\Delta = VWV^\dagger W^\dagger$ , $V_0$ and $W_0$ (0-level approximations of $V$ and $W$ ) be found through exhaustive searches; Then $U_1 = V_0 W_0 V_0^\dagger W_0^\dagger U_0$.

(iii) 2-level approximation $U_2$: setting $\Delta' = UU_1^\dagger$ and deriving $V'$ and $W'$ by GC-decomposition for $\Delta'$; Computing 1-level approximations $V_1'$ and $W_1'$ (1-level approximations of $V'$ and $W'$ ) using recursive SKA calls; $U_2 = V_1' W_1' (V_1')^\dagger (W_1')^\dagger U_1$ be assembled.

(iv) Higher-level approximations $U_n$ are iteratively generated through successive GC-decompositions, achieving exponential precision scaling.

The GC-decomposition constitutes the algorithmic core, factorizing the residual operator $\Delta$ into $VWV^{\dagger}W^{\dagger}$. This non-trivial factorization depends on solving the equation $\sin(\theta/2) = 2\sin^2(\phi/2)\sqrt{1-\sin^4(\phi/2)}$ and the definitions of the $V$- and $W$-matrices. Comprehensive technical details of SKA implementation can be found in foundational works [40].

The conventional SKA employs exhaustive search for 0-level approximations, incurring inherent limitations such as exponential resource scaling that imposes strict constraints on maximum braid length and incurs prohibitive computational overhead. Our GA implementation strategically circumvents these limitations through stochastic optimization.

The quantum compiling framework maps naturally to GA components as shown in Table I.

**Table I.** The mapping between genetic algorithm and quantum compiling.

| **Genetic algorithm components** | **Quantum compiling elements** |
| --- | --- |
| Individuals | Candidate braidwords. |
| Population | Ensemble of $\sim 10^3$ braidwords configurations. |
| Mutation | Several EBMs modifications in a braidword. |
| Crossover | Hybridization of two parent braidwords. |
| Fitness function | Global-phase-invariant distance metric. |

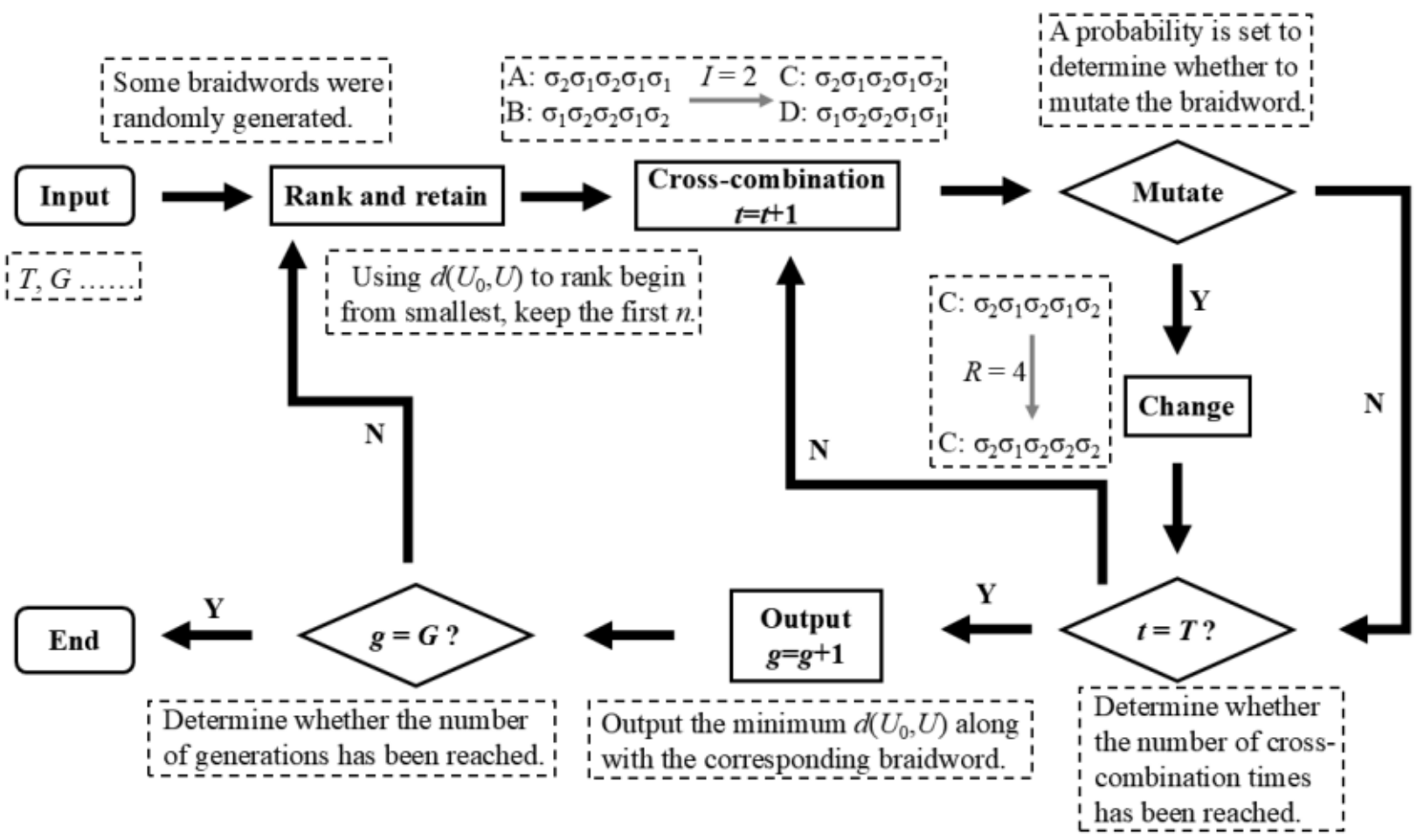

**Fig. 2**: Flowchart of GA.

Fig. 2 presents the workflow of GA, which operates through the following key stages:

① **Initialization**:

According to the specified length L (referring to the number of EBMs in a braid word), thousands of valid braid words (where a valid braid word contains no adjacent $\{\sigma_1\sigma_1^{-1}, \sigma_1^{-1}\sigma_1, \sigma_2\sigma_2^{-1}, \sigma_2^{-1}\sigma_2\}$ pairs) are randomly generated as the initial population.

② **Evolutionary Operations**:

Crossover: Randomly select two parent braidwords for sequence hybridization in population. (e.g., given two selected braidwords $A(\sigma_2\sigma_1\sigma_2\sigma_1\sigma_1)$ and $B(\sigma_1\sigma_2\sigma_2\sigma_1\sigma_2)$, a random integer $I$ within the range $0 \le I \le \mathrm{L}$ is chosen. If $I = 2$, both braidwords are truncated after the second EBM, and the truncated segments are then swapped via crossover to generate two new braid words $C(\sigma_2\sigma_1\sigma_2\sigma_1\sigma_2)$ and $D(\sigma_1\sigma_2\sigma_2\sigma_1\sigma_1)$.)

Mutation: Apply EBM substitutions to offspring with probability $p$. (e.g., if a braidword $C(\sigma_2\sigma_1\sigma_2\sigma_1\sigma_2)$ is selected to undergo mutation, a random integer $R$ within the range $0 \le R \le L$ is chosen. If $R = 4$, the fourth EBM in the braidword will be replaced with another EBM from the available set, resulting in a modified braidword $C(\sigma_2\sigma_1\sigma_2\sigma_2\sigma_2)$.)

Then the new braidwords generated through crossover and mutation operations are recorded into a new population, and the times of crossover increase by one (that is, t=t+1).

③ **Convergence Check**:

If t < $T$: Return to Step ② ($T$ represents the preset total number of crossover operations to be performed.).

Else: The optimal braidword exhibiting minimal distance to the target qubit gate be output, then the algorithm enters the next crossover cycle Step ④.

④ **Population Update**:

The population obtained from crossover in the previous generation undergoes truncation selection, where the best-performing offspring (i.e., the top n genomes with the smallest d-values) are retained to form a new population for the next generation. The iterative process repeats Steps ②-③ with the refined population. Then the generation increase by one (that is, g= g+1).

⑤ **Termination Criterion**:

If g ≤ $G$: Continue to Step ④ (G represents the preset number of generations to be completed.).

Else: Terminate the algorithm.

For implementation specifics of this GA-enhanced SKA framework, including hyperparameter tuning ($p$, $T$, $N$, $G$, etc), see the Reference [46].

Makhlin characterized two-qubit gates through real-valued local invariants [48], while Zhang et al. established a geometric framework for two-qubit operations by integrating these invariants with the SU(4) Cartan decomposition [49]. Two gates belong to the same local equivalence class if they can be made identically through one-qubit operations. Specifically, if two two-qubit gates $U$ and $V$ satisfy $U = (k_1 \otimes k_2) \bullet V \bullet (k_3 \otimes k_4)$, where each $k_i$ represents a one-qubit operation, then they belong to the same local equivalence class. The essence of this classification lies in stripping away the local operations that act solely on individual qubits, thereby focusing on the genuine non-local entanglement characteristics of the two-qubit gates. Through the geometric framework of Makhlin invariants and the SU(4) Cartan decomposition, local equivalence classes are mapped to points in a three-dimensional parameter space. This transforms the complex problem of gate compilation into a simplified search within this lower-dimensional space. Such an approach not only significantly reduces the optimization dimensionality but also expands the effective range for approximating target gates. Approximating a local equivalence class of two-qubit proves significantly simpler than direct gate synthesis due to reduced constraints [50]. The two-qubit EBMs based on Fibonacci anyons has been precisely solved by Cui et al. [51]. Recent work by Burke et al. demonstrated high-fidelity approximations of the local equivalence class [CNOT] using Fibonacci anyon-based two-qubit EBMs [28].

We briefly outline the protocol for determining a local equivalence class:

Let $B$ denote the braidword matrix. Through the direct sum decomposition: $B = M \oplus A$, $M$ is the non-computational sector, $A$-matrix corresponds to computational subspace. The target gate $U$ ($A$-matrix or a standard two-qubit gate) is then transformed into the Bell basis via

$$U_B = Q^{\dagger} U Q, Q = \frac{1}{\sqrt{2}} \begin{bmatrix} 1 & 0 & 0 & i \\ 0 & i & 1 & 0 \\ 0 & i & -1 & 0 \\ 1 & 0 & 0 & -i \end{bmatrix} \tag{4}$$

where $Q$ is the Bell basis transformation matrix.

A complete set of real-valued local invariants is computationally determined through the following formulas:

$$g_1 = \mathrm{Re}\left\{ \frac{tr^2(m_U)}{16 \cdot \det(U)} \right\}, g_2 = \mathrm{Im}\left\{ \frac{tr^2(m_U)}{16 \cdot \det(U)} \right\}, g_3 = \frac{tr^2(m_U) - tr(m_U^2)}{4 \cdot \det(U)}, m_U = U_B^{\mathrm{T}} U_B \tag{5}$$

The local invariant [CNOT] can be calculated by the above three formulas:

$$g_1(CNOT) = 0, g_2(CNOT) = 0, g_3(CNOT) = 1$$

The formula for measuring the distance between $A$-matrix in the braidword and the local invariant [CNOT] is:

$$d^{CNOT}(A) = \sum_{i=1}^{3} \Delta g_i^2, \Delta g_i = \left| g_i(A) - g_i(CNOT) \right| \tag{6}$$

To mitigate leakage errors, the approximate unitary of $M$-value and $A$-matrix must be enforced. The unitary of these matrices is quantified through the following metric:

$$M_{11} = \sqrt{M^* M}, d^U = Tr(\sqrt{a^\dagger a}), a = A^\dagger A - \mathrm{I}, \tag{7}$$

where I is a four-dimensional identity matrix.

The EBMs of $SU(2)_{5,6,7}$ two-qubit systems were employed to approximate the local equivalence class [CNOT], with exhaustive search adopted for shorter braid lengths and GA implemented for extended configurations.

## 3 Results and discussions

### 3.1 The construction of one-qubit gate

Under identical parameter configurations, the results computed via the GA-enhanced SKA method for $SU(2)_{3,5,6,7}$ anyon models are shown in Fig. 3, where $SU(2)_3$ corresponds to the well-known Fibonacci anyon model. While the SKA framework achieves exponential reduction in computational distance, the number of GA searches required for 0-level approximations triples with each recursion level. Consequently, higher-level approximations (e.g., 4-level) incur prohibitive computational costs—81 GA searches are required for 4-level approximations. Remarkably, 3-level approximations attain precision sufficient for practical quantum computing, thus our analysis is restricted to this level. Both $H$- and $T$-gates are successfully constructed across all models, with $SU(2)_7$ demonstrating superior performance among the four. At 3-level approximation, $SU(2)_7$ achieves gate errors on the order of $10^{-6}$ for both standard $H$- and $T$-gates. Fig. 3(a) presents the $H$-gate compilation results. The fidelity of $SU(2)_{3,5,6}$ implementations is comparable but slightly inferior to that of $SU(2)_7$. As shown in Fig. 3(b), $SU(2)_{5,7}$ exhibit nearly equivalent $T$-gate precision, outperforming $SU(2)_{3,6}$. Notably, $SU(2)_{5,7}$ achieve $10^{-5}$ errors at 2-level approximation—matching the 3-level precision of $SU(2)_{3,6}$. Since $10^{-5}$ errors are fully compatible with quantum computing requirements, $SU(2)_{5,7}$ implementations significantly reduce redundant braiding operations compared to $SU(2)_{3,6}$—achieving equivalent precision with fewer approximation levels (2-level vs 3-level approximation, corresponding to $30\times5^2$ vs $30\times5^3$ braiding times).

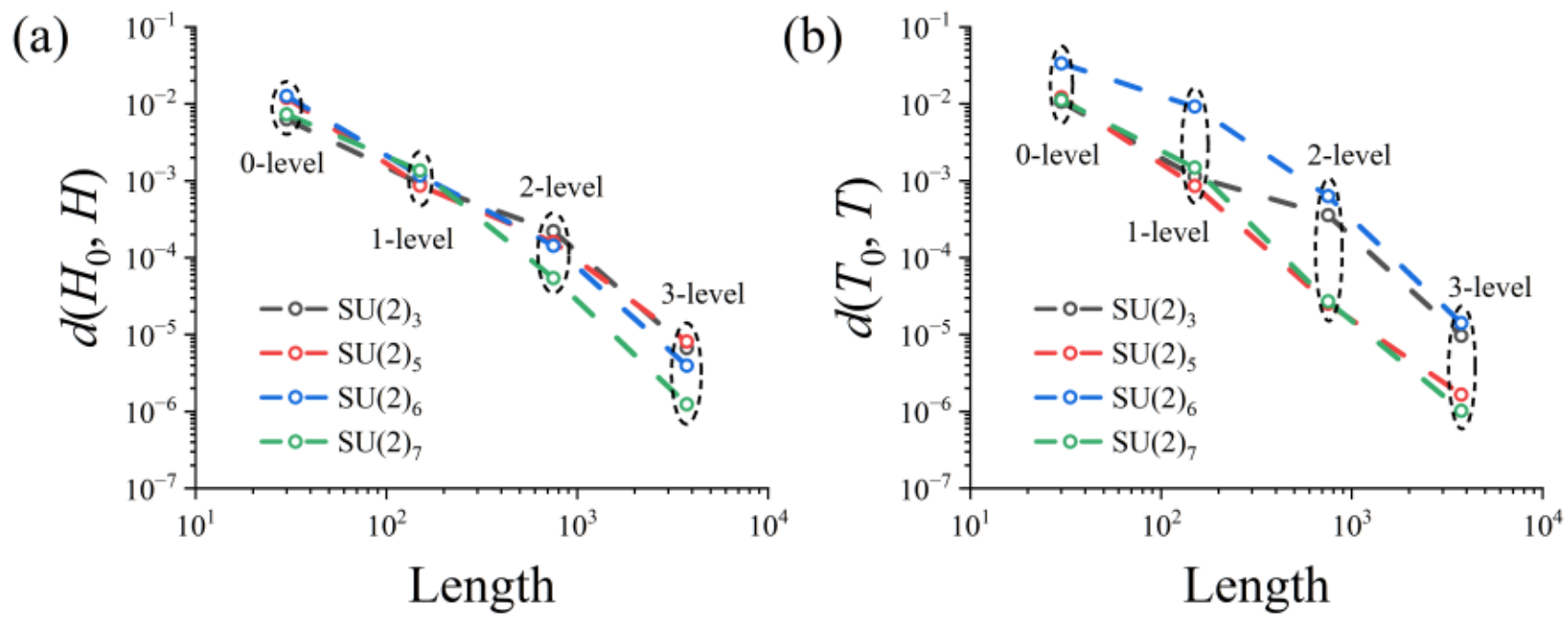

**Fig 3**: One-qubit gate compilation via GA-enhanced SKA for $SU(2)_{3,5,6,7}$ anyon models. Basic braid length $l_0$=30. (a) *H*-gate. (b) *T*-gate.

Table II catalogs the 0-level braidwords and corresponding $d(U_0,U)$ for *H*-/*T*-gate approximations across $SU(2)_{3,5,6,7}$ models. While $SU(2)_3$ achieves exceptionally low $d(U_0,U)$ for *H*-gate at 0-level, $SU(2)_7$ demonstrates superior performance at higher approximation levels. This counterintuitive result arises from the GC-decomposition: $SU(2)_7$ consistently yields lower $d(V_0, V)$ and $d(W_0, W)$ for GC components *V* and *W*. The enhanced higher-level fidelity stems from balanced error suppression across all GC-decomposition (Perform approximate optimization on *V* and *W* obtained from each GC-decomposition), stages rather than isolated 0-level optimization of *U*. From another perspective, the EBMs of the $SU(2)_k$ model depend on the parameter $q = e^{i\frac{2\pi}{k+2}}$. As k increases, the phase of *q* becomes more finely spaced, implying that the braid group representations become denser and more effectively cover the unitary group. This could potentially lead to improved approximation accuracy. Nonetheless, the underlying reasons why $SU(2)_7$ enables more precise construction of *H*- and *T*-gates remain an open question worthy of further exploration.

**Table II.** 0-level braidwords and $d(U_0,U)$ metrics for *H*-/*T*-gates. A/B/C/D corresponding to $\sigma_1/\sigma_2/\sigma_1^{-1}/\sigma_2^{-1}$. The world-lines of the anyons corresponding to these braidwords are presented in Appendix E.

| | Models | Braidwords | $d(U_0, U)$ |
|---|---|---|---|
| *H*-gate | $SU(2)_3$ | CDADDADCBADDADDDDCDADADADADADD | 0.00626791 |
| | $SU(2)_5$ | ADCCDCDABBADCCDABBBADADDDAAAAA | 0.01197934 |
| | $SU(2)_6$ | BBBBCCBBBCBBBCBABBABBCBBBBABBB | 0.01265547 |
| | $SU(2)_7$ | DCBADDCBBCDDCDCBCBAAADAAAAADDC | 0.00730905 |
| *T*-gate | $SU(2)_3$ | ADDDCDDADDADADCDCDADDADDDDDCCD | 0.01063365 |
| | $SU(2)_5$ | CDCDDCDDAADDCCBBBADADDADADAABC | 0.01211672 |
| | $SU(2)_6$ | ABBBBABBBCCBBCBBAABBBBBBABBBBB | 0.03361571 |
| | $SU(2)_7$ | DCCCCDDCBAAAADCBCBBBBBCBBCDCDD | 0.01121239 |

### 3.2 The construction of two-qubit gate

Computational analysis confirms that $SU(2)_{5,6,7}$ two-qubit EBMs successfully approximate the local equivalence class [CNOT], requiring only about thirty braiding operations to achieve ultra-low errors ($<10^{-6}$) comparable to Fibonacci anyon implementations. Note that the data of $SU(2)_3$ comes from the references [28], we have verified the correctness of the data. Fig. 4(a) displays compilation results without inverse EBMs optimization. The vertical dashed line demarcates methodology regimes: exhaustive search for lengths ≤13 (left) and GA implementations for lengths >13 (right), where combinatorial complexity prohibits brute-force approach. All models achieve the local equivalence class [CNOT] distances $<10^{-6}$ at lengths ≥31. Remarkably, $SU(2)_{6,7}$ systems yield braidword with equivalence errors $<10^{-10}$ at specific length 31. Unitary requirements ($|M_{11}|\approx 1$, $d^U \approx 0$) are satisfied approximately: Fig. 4(b) show the $|M_{11}| > 0.94$ and Fig. 4(c) show the $d^U < 0.1$ across all lengths.

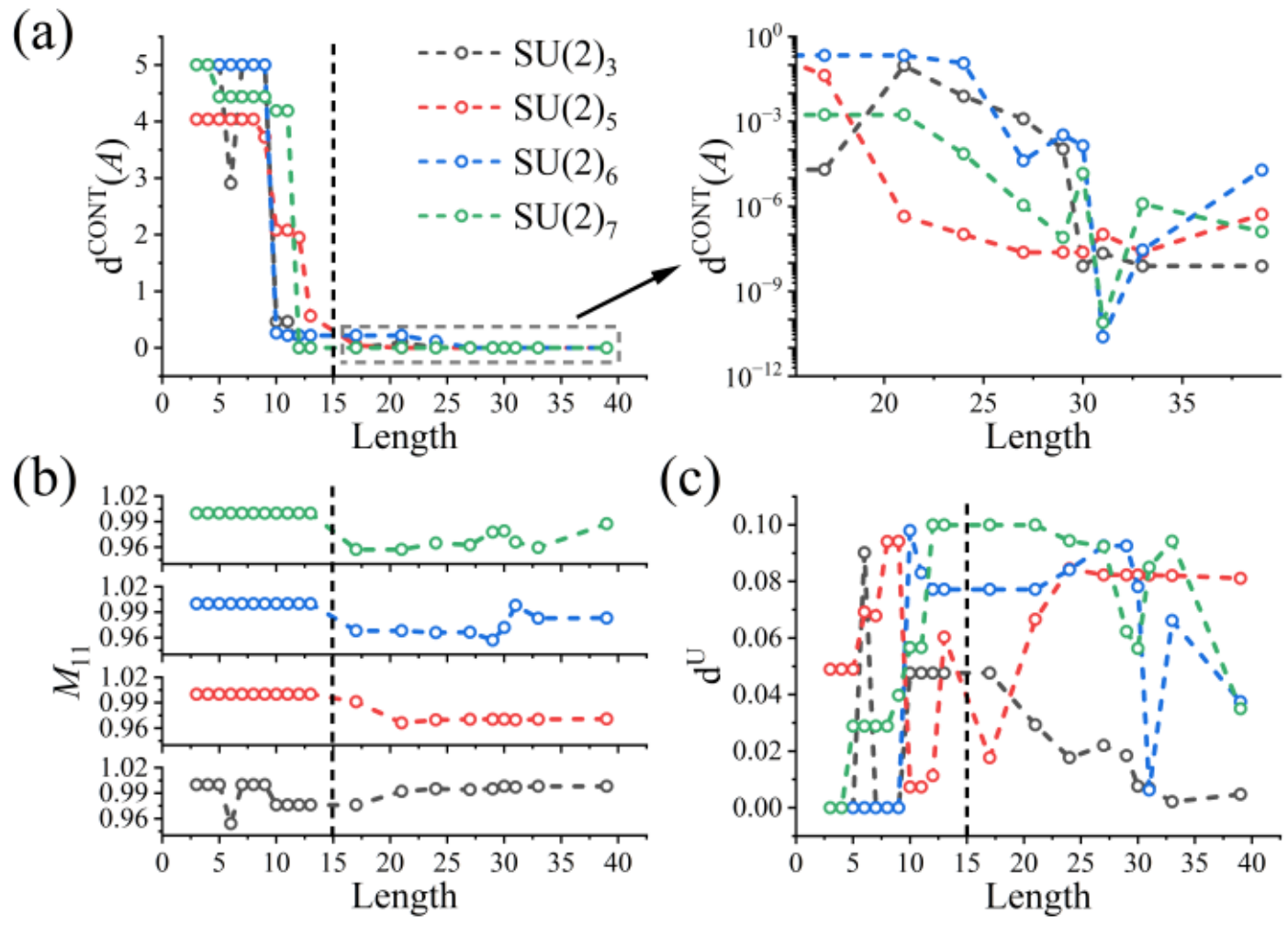

**Fig. 4**: Approximation of the local equivalence class [CNOT] using $SU(2)_{3,5,6,7}$ anyon models. The vertical dashed line demarcates the methodological transition between exhaustive search and GA implementations, with no inverse matrix of EBMs. (a) Compilation fidelity: local equivalence class [CNOT] distances as a function of braid length. (b) Non-computational sector unitary $|M_{11}|$ as a function of braid length (c) Computational subspace $A$-matrix unitarity $d^U$ as a function of braid length.

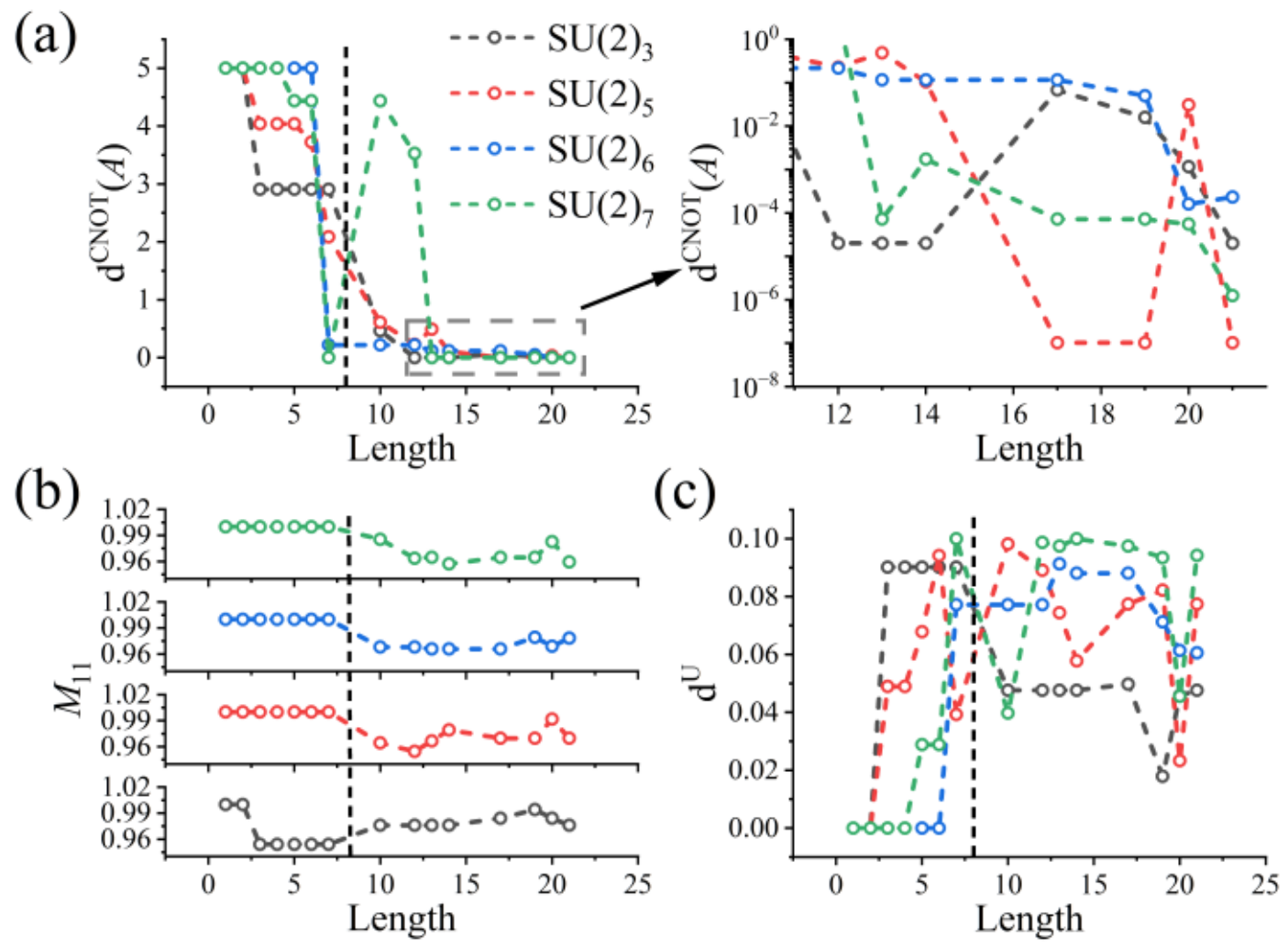

**Fig. 5**: Approximation of the local equivalence class [CNOT] using $SU(2)_{3,5,6,7}$ anyon models. The vertical dashed line demarcates the methodological transition between exhaustive search and GA implementations, adding the inverse matrix of EBMs. (a) Compilation fidelity: local equivalence class [CNOT] distances as a function of braid length. (b) Non-computational sector unitary $|M_{11}|$ as a function of braid length (c) Computational subspace $A$-matrix unitarity $d^{U}$ as a function of braid length.

Fig. 5(a) displays compilation results with inverse EBMs integration. The inclusion of inverse matrices increases braidword diversity from 5 to 10 EBM types, restricting exhaustive search feasibility to lengths ≤7 (left of dashed line). GA implementations address lengths >7 (right of dashed line). Inverse matrix optimization yields no statistically significant $d^{CNOT}(A)$ reduction. Only $SU(2)_5$ achieves sub-$10^{-6}$ $d^{CNOT}(A)$ within length 21. Fig 5(b)-(c) confirm preserved unitarity under inverse EBM conditions: $|M_{11}| > 0.94$ and $d^{U} < 0.1$, consistent with no-inverse implementations. Comparative analysis of Fig. 4(a) and 5(a) demonstrates that braid length extension surpasses inverse matrix optimization for $d^{CNOT}(A)$ minimization. Table IV enumerates the minimal- $d^{CNOT}(A)$ braidword across all investigated lengths. The low-error approximation of the local equivalence class [CNOT] using braidword composed of EBMs from $SU(2)_{3,5,6,7}$ demonstrates the capability of these models to construct entangling gates.

Table III. The minimal-$d^{CNOT}(A)$ and corresponding braidword across all investigated lengths. A/B/C/D/E/F/G/H/I/J corresponding to $\sigma_1 / \sigma_2 / \sigma_3 / \sigma_4 / \sigma_5 / \sigma_1^{-1} / \sigma_2^{-1} / \sigma_3^{-1} / \sigma_4^{-1} / \sigma_5^{-1}$. The world-lines of the anyons corresponding to these braidwords are presented in Appendix E.

| | Moldes | Braidwords | $d^{CNOT}(A)$ |
|---|---|---|---|
| Adding the inverse matrix | $SU(2)_3$ | CAIJCDDCJIJC | $2.00\times10^{-05}$ |
| | $SU(2)_5$ | HHHHEHHHHDJDHHHHH | $1.02\times10^{-07}$ |
| | $SU(2)_6$ | GCFCAIJDJICIJBJIGECC | $1.60\times10^{-04}$ |
| | $SU(2)_7$ | ICJCACCCDEEBDCCGAGCAC | $1.24\times10^{-06}$ |
| No | $SU(2)_3$ | CCCCCDAEDECCECCACEAEDDDDDCAAAD | $7.78\times10^{-09}$ |
| inverse | $SU(2)_5$ | CCCBBCCCCCCBAABACCCCCCCCCAE | $2.37\times10^{-08}$ |
| matrix | $SU(2)_6$ | DDBCBEBBCECCDDBAAEDBDCACCBBCBCB | $2.41\times10^{-11}$ |
| | $SU(2)_7$ | CCCCCDADDADDEDECDADDCCAEDEDCCCD | $7.81\times10^{-11}$ |

The Fibonacci anyon model admits an exact implementation of the local equivalence class [SWAP] at braid length 9[28].

The local equivalence class [SWAP] are computed through the following protocol:

$$g_1(SWAP) = -1, \quad g_2(SWAP) = 0, \quad g_3(SWAP) = -3,$$

The corresponding distance formula becomes:

$$d^{SWAP}(A) = \sum_{i=1}^{3} \Delta g_i^2, \quad \Delta g_i = \left| g_i(A) - g_i(SWAP) \right|. \tag{8}$$

Computational analysis reveals that $SU(2)_{5,6,7}$ models each admit exact implementations ($d^{SWAP} \sim 10^{-14}$) of the local equivalence class [SWAP] at braid length 9. The corresponding optimal braidwords are cataloged in Table III. Brute-force search across $SU(2)_{5,6,7}$ systems consistently yields braidwords with: $d^{CNOT}(A) = 0$, $|M_{11}| = 1$ and $d^{SWAP} \approx 0$. It can be guessed the native SWAP gate realizability in $SU(2)_k$ (k=3, and k≥5) anyon systems through minimal 9-step braiding operations with no inverse matrix of EBMs.

**Table**IV**:** Braidwords achieving 0 distance to the local equivalence class **[SWAP]** at length 9 for $SU(2)_{5,6,7}$ anyon models. A/B/C/D/E corresponding to $\sigma_1/\sigma_2/\sigma_3/\sigma_4/\sigma_5$. The world-lines of the anyons corresponding to these braidwords are presented in Appendix E.

| Models | Braidwords | $d^{swap}$ | $M_{11}$ | $d^{SWAP}$ |
|---|---|---|---|---|
| $SU(2)_5$ | CDBACEBDC | $1.48\times10^{-32}$ | 1 | $5.88\times10^{-15}$ |
| $SU(2)_6$ | CDEBCADBC | $1.23\times10^{-32}$ | 1 | $1.31\times10^{-14}$ |
| $SU(2)_7$ | CBADCBEDC | $1.23\times10^{-32}$ | 1 | $5.82\times10^{-15}$ |

## 4 Conclusions

In summary, the one- and two-qubit EBMs for $SU(2)_k$ anyon models — encoded via spatially arranged configurations of 3 or 6 topological spin-1/2 anyons — are analytically determined using *F*- and *R*-symbols derived from the *q*-deformed SU(2) representation theory. For exemplar cases (k=5, 6, 7), we construct a universal quantum gate set {*H*-gate, *T*-gate, CNOT-gate} using $SU(2)_{5,6,7}$ EBMs. The GA-enhanced SKA synthesizes *H*- and *T*-gates with 3-level approximation errors $d(U_0,U)$ of $10^{-5}$–$10^{-6}$ magnitude. The local equivalence class [CNOT] is accurately approximated by braidwords composed of these EBMs (up to one-qubit operations), achieving local equivalence class distances $\leq 10^{-8}$ at braid lengths ~30. These ultralow errors satisfy fault-tolerant quantum computation thresholds [52-57], providing numerical verification of $SU(2)_k$ anyon models' capacity for universal quantum computation. Finally, exact implementations of the local equivalence class [SWAP] are achieved through 9-step braiding operations in $SU(2)_{5,6,7}$ systems, we can guess the native realizability of SWAP gates in $SU(2)_k$ (k=3, and k≥5) anyon architectures.

**Funding** This work is supported by the National Natural Science Foundation of China (Grant Nos. 12374046, 11204261), College of Physics and Optoelectronic Engineering training program, a Key Project of the Education Department of Hunan Province (Grant No. 19A471), Natural Science Foundation of Hunan Province (Grant No. 2018JJ2381), Shanghai Science and Technology Innovation Action Plan (Grant No. 24LZ1400800). Education Department of Hunan Province (Grant No. 24C0316)

**Acknowledgements** We are grateful to Xuyang Huang (Shanghai University) for the discussion on the code implementation of the genetic algorithm. We sincerely thank the anonymous reviewers for the valuable comments and suggestions, which helped us significantly improve this paper.

**Data Availability Statement** The datasets generated during and/or analyzed during the current study are available from the corresponding author on reasonable request

**Declarations**

**Conflict of interest** No potential conflict of interest was reported by the authors. All authors of this manuscript have read and approved the final version submitted, and contents of this manuscript have not been copyrighted or published previously and are not under consideration for publication elsewhere.

**Appendix A The q-deformed representation theory of SU(2)**

"q-integers" are definded by $\left[n\right]_q \equiv \frac{q^{n/2}-q^{-n/2}}{q^{1/2}-q^{-1/2}}$ , where the deformation parameter $q=e^{i\frac{2\pi}{k+2}}$ (k is an integer for $SU(2)_k$).

The $R$-symbol, corresponds to the rotation of anyons with topological spins $j_1$ and $j_2$, resulting in an anyon with topological spin $j$, defined by :

$$R_j^{j_1,j_2} = (-1)^{j-j_1-j_2} q^{\frac{1}{2}[j(j+1)-j_1(j_1+1)-j_2(j_2+1)]}$$

The $F$-symbol is defined by the following formula,

$$\left[F_j^{j_1,j_2,j_3}\right]_{j_{12},j_{23}} = (-1)^{j_1+j_2+j_3+j} \sqrt{[2j_{12}+1]_q [2j_{23}+1]_q} \begin{Bmatrix} j_1 & j_2 & j_{12} \\ j_3 & j & j_{23} \end{Bmatrix}_q ,$$

where $j_1, j_2$, and $j_3$ are the topological spins of the initial anyons. These anyons fuse into a final anyon with topological spin $j$, mediated by an intermediate fusion channel that transitions from an anyon with topological spin $j_{12}$ to one with topological spin $j_{23}$. And

$$\begin{Bmatrix} j_1 & j_2 & j_{12} \\ j_3 & j & j_{23} \end{Bmatrix}_q = \Delta(j_1,j_2,j_{12})\Delta(j_{12},j_3,j)\Delta(j_2,j_3,j_{23})\Delta(j_1,j_{23},j)$$

$$\times \sum_z \left\{ \begin{array}{l} \dfrac{(-1)^z [z+1]_q!}{[z-j_1-j_2-j_{12}]_q![z-j_{12}-j_3-j]_q![z-j_2-j_3-j_{23}]_q![z-j_1-j_{23}-j]_q!} \\ \times \dfrac{1}{[j_1+j_2+j_3+j-z]_q![j_1+j_{12}+j_3+j_{23}-z]_q![j_2+j_{12}+j+j_{23}-z]_q!} \end{array} \right\},$$

$$\Delta(j_1,j_2,j_3) = \sqrt{\frac{[-j_1+j_2+j_3]_q![j_1-j_2+j_3]_q![j_1+j_2-j_3]_q!}{[j_1+j_2+j_3+1]_q!}}, \quad [n]_q! \equiv \prod_{m=1}^{n}[m]_q ,$$

$$z_{\min} \le z \le z_{\max} \left( \begin{array}{l} z \in \text{integer}, \\ z_{\min} \in \max\{j_1+j_2+j_{12},\, j_{12}+j_3+j,\, j_2+j_3+j_{23},\, j_1+j_{23}+j\}, \\ z_{\max} \in \min\{j_1+j_2+j_3+j,\, j_1+j_{12}+j_3+j_{23},\, j_2+j_{12}+j+j_{23}\} \end{array} \right),$$

the range of $z$ is determined by $n \ge 0$ in $[n]_q!$.

## Appendix B: *F*-Matrices and *R*-Symbols for calculating EBMs of SU(2)$_{5,6,7}$ anyon models

The definitions of the F- and R-symbols are given in the following figure:

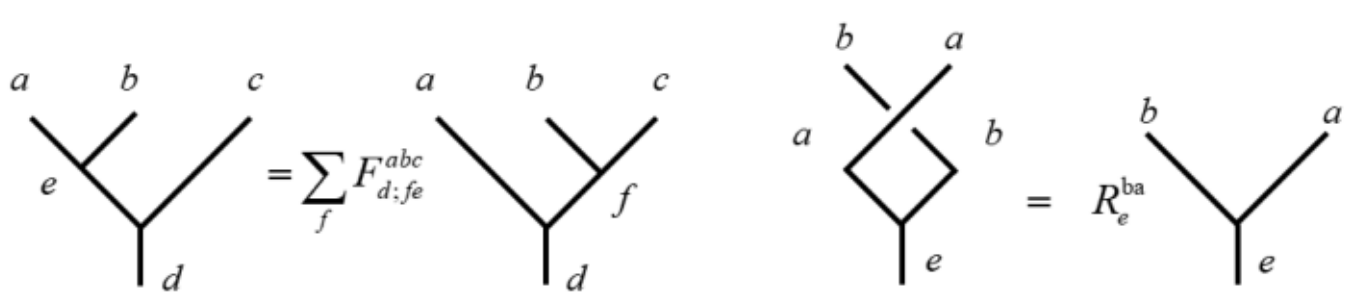

Definition of $F^{abc}_{d;fe}$ Definition of $R^{ab}_e$

The explicit definitions of the $F$-symbols and $R$-symbols used to compute EBMs in SU(2)$_{5,6,7}$ anyon models are provided below:

The explicit $F$-matrices and $R$-symbols required for computing one-qubit and two-qubit EBMs in $SU(2)_{5,6,7}$ anyon models are listed below. These $F$-matrices and $R$-symbols are analytically derived from the $q$-deformed representation theory of SU(2).

$$\mathrm{SU}(2)_5:$$
$$R_0^{11} = \text{-0.78183148+0.62348980j} \quad R_2^{11} = \text{0.97492791+0.22252093j}$$
$$F_1^{111} = \begin{bmatrix} F_{1;00}^{111} & F_{1;02}^{111} \\ F_{1;20}^{111} & F_{1;22}^{111} \end{bmatrix} = \begin{bmatrix} \text{-0.55495813} & 0.83187828 \\ 0.83187828 & 0.55495813 \end{bmatrix}$$
$$F_2^{112} = \begin{bmatrix} F_{2;10}^{112} & F_{2;12}^{112} \\ F_{2;30}^{112} & F_{2;32}^{112} \end{bmatrix} = \begin{bmatrix} -0.66711458 & 0.74495512 \\ 0.74495512 & 0.66711458 \end{bmatrix}$$

$$\mathrm{SU}(2)_6:$$
$$R_0^{11} = \text{-0.83146961+0.55557023j} \quad R_2^{11} = \text{0.98078528+0.19509032j}$$
$$F_1^{111} = \begin{bmatrix} F_{1;00}^{111} & F_{1;02}^{111} \\ F_{1;20}^{111} & F_{1;22}^{111} \end{bmatrix} = \begin{bmatrix} \text{-0.54119610} & 0.84089642 \\ 0.84089642 & 0.54119610 \end{bmatrix}$$
$$F_2^{112} = \begin{bmatrix} F_{2;10}^{112} & F_{2;12}^{112} \\ F_{2;30}^{112} & F_{2;32}^{112} \end{bmatrix} = \begin{bmatrix} \text{-0.64359425} & 0.76536686 \\ 0.76536686 & 0.64359425 \end{bmatrix}$$

$$\mathrm{SU}(2)_7:$$
$$R_0^{11} = \text{-0.86602540+0.5j} \quad R_2^{11} = \text{0.98480775+0.17364818j}$$
$$F_1^{111} = \begin{bmatrix} F_{1;00}^{111} & F_{1;02}^{111} \\ F_{1;20}^{111} & F_{1;22}^{111} \end{bmatrix} = \begin{bmatrix} \text{-0.53208889} & 0.84668850 \\ 0.84668850 & 0.53208889 \end{bmatrix}$$
$$F_2^{112} = \begin{bmatrix} F_{2;10}^{112} & F_{2;12}^{112} \\ F_{2;30}^{112} & F_{2;32}^{112} \end{bmatrix} = \begin{bmatrix} -0.62843523 & 0.77786191 \\ 0.77786191 & 0.62843523 \end{bmatrix}$$

## Appendix C: General calculation process for EBMs in $SU(2)_{5,6,7}$ anyon models

The braiding operator $\sigma_1^{(3)}$ is straightforward to construct: it involves braiding the first and second anyons within a triad of spatially arranged non-Abelian quasiparticles, with the $R$-symbols encoded into the diagonal entries of the corresponding unitary matrix:

$$\sigma_1^{(3)} = \begin{bmatrix} R_0^{11} & 0 \\ 0 & R_2^{11} \end{bmatrix}$$

The braiding operator $\sigma_2^{(3)}$ is constructed through the following sequence of topological operations:

1. Basis transformation: Applying an inverse $F$-move to modify the fusion basis ordering;
2. Anyon braiding: Performing an $R$-move to braid the second and third anyons;
3. Basis restoration: Return to the original fusion basis via the $F$-move.

This process is given by the following figure:

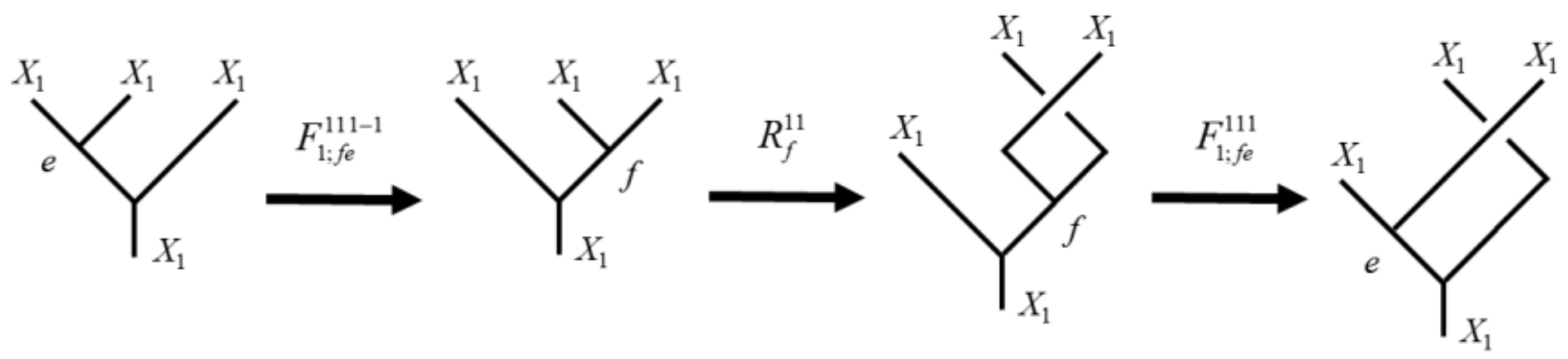

So $\sigma_2^{(3)} = F_1^{111-1} R^{11} F_1^{111} = \begin{bmatrix} F_{1;00}^{111} R_0^{11} F_{1;00}^{111} + F_{1;20}^{111} R_2^{11} F_{1;20}^{111} & F_{1;00}^{111} R_0^{11} F_{1;02}^{111} + F_{1;20}^{111} R_2^{11} F_{1;22}^{111} \\ F_{1;02}^{111} R_0^{11} F_{1;00}^{111} + F_{1;22}^{111} R_2^{11} F_{1;20}^{111} & F_{1;02}^{111} R_0^{11} F_{1;02}^{111} + F_{1;22}^{111} R_2^{11} F_{1;22}^{111} \end{bmatrix}$.

The braiding operators $\sigma_1^{(6)} / \sigma_5^{(6)}$ in a 6-anyon two-qubit system are constructed by braiding the first and second anyons $\sigma_1^{(6)}$ or the fifth and sixth anyons $\sigma_5^{(6)}$ without requiring $F$-moves. The corresponding $R$-symbols are directly encoded into the diagonal entries of a 5-dimensional unitary matrix, yielding:

$$\sigma_1^{(6)} = \begin{bmatrix} R_2^{11} & 0 & 0 & 0 & 0 \\ 0 & R_0^{11} & 0 & 0 & 0 \\ 0 & 0 & R_0^{11} & 0 & 0 \\ 0 & 0 & 0 & R_2^{11} & 0 \\ 0 & 0 & 0 & 0 & R_2^{11} \end{bmatrix}, \quad \sigma_5^{(6)} = \begin{bmatrix} R_2^{11} & 0 & 0 & 0 & 0 \\ 0 & R_0^{11} & 0 & 0 & 0 \\ 0 & 0 & R_2^{11} & 0 & 0 \\ 0 & 0 & 0 & R_0^{11} & 0 \\ 0 & 0 & 0 & 0 & R_2^{11} \end{bmatrix}.$$

The braiding operators $\sigma_2^{(6)} / \sigma_4^{(6)}$ in a six-anyon two-qubit system are constructed by braiding the second and third anyons (for $\sigma_2^{(6)}$) or the fourth and fifth anyons (for $\sigma_4^{(6)}$). To achieve this, $F$-moves are applied to modify the fusion basis ordering, followed by performing the corresponding braiding operations on each basis state. The effects of these operations on all basis states are then compose into the unitary matrix representation of the operators, yielding the corresponding braiding matrices:

$$\sigma_2^{(6)} = \begin{bmatrix} R_2^{11} & 0 & 0 & 0 & 0 \\ 0 & F_{1;00}^{111} R_0^{11} F_{1;00}^{111} + F_{1;20}^{111} R_2^{11} F_{1;20}^{111} & 0 & F_{1;00}^{111} R_0^{11} F_{1;02}^{111} + F_{1;20}^{111} R_2^{11} F_{1;22}^{111} & 0 \\ 0 & 0 & F_{1;00}^{111} R_0^{11} F_{1;00}^{111} + F_{1;20}^{111} R_2^{11} F_{1;20}^{111} & 0 & F_{1;00}^{111} R_0^{11} F_{1;02}^{111} + F_{1;20}^{111} R_2^{11} F_{1;22}^{111} \\ 0 & F_{1;02}^{111} R_0^{11} F_{1;00}^{111} + F_{1;22}^{111} R_2^{11} F_{1;20}^{111} & 0 & F_{1;02}^{111} R_0^{11} F_{1;02}^{111} + F_{1;22}^{111} R_2^{11} F_{1;22}^{111} & 0 \\ 0 & 0 & F_{1;02}^{111} R_0^{11} F_{1;00}^{111} + F_{1;22}^{111} R_2^{11} F_{1;20}^{111} & 0 & F_{1;02}^{111} R_0^{11} F_{1;02}^{111} + F_{1;22}^{111} R_2^{11} F_{1;22}^{111} \end{bmatrix},$$

$$\sigma_4^{(6)} = \begin{bmatrix} R_2^{11} & 0 & 0 & 0 & 0 \\ 0 & F_{1;00}^{111} R_0^{11} F_{1;00}^{111} + F_{1;20}^{111} R_2^{11} F_{1;20}^{111} & F_{1;00}^{111} R_0^{11} F_{1;02}^{111} + F_{1;20}^{111} R_2^{11} F_{1;22}^{111} & 0 & 0 \\ 0 & F_{1;02}^{111} R_0^{11} F_{1;00}^{111} + F_{1;22}^{111} R_2^{11} F_{1;20}^{111} & F_{1;02}^{111} R_0^{11} F_{1;02}^{111} + F_{1;22}^{111} R_2^{11} F_{1;22}^{111} & 0 & 0 \\ 0 & 0 & 0 & F_{1;00}^{111} R_0^{11} F_{1;00}^{111} + F_{1;20}^{111} R_2^{11} F_{1;20}^{111} & F_{1;00}^{111} R_0^{11} F_{1;02}^{111} + F_{1;20}^{111} R_2^{11} F_{1;22}^{111} \\ 0 & 0 & 0 & F_{1;02}^{111} R_0^{11} F_{1;00}^{111} + F_{1;22}^{111} R_2^{11} F_{1;20}^{111} & F_{1;02}^{111} R_0^{11} F_{1;02}^{111} + F_{1;22}^{111} R_2^{11} F_{1;22}^{111} \end{bmatrix}.$$

The direct product relationship between $\sigma_1^{(6)} / \sigma_2^{(6)} / \sigma_4^{(6)} / \sigma_5^{(6)}$ and $\sigma_1^{(3)} / \sigma_2^{(3)}$ can be easily discovered.

The braiding operator $\sigma_3^{(6)}$, acting on the third and fourth anyons within a 6-anyon two-qubit system, is constructed by sequentially applying $F$-moves and $R$-moves to each basis state of the fusion space. As illustrated in follow figures:

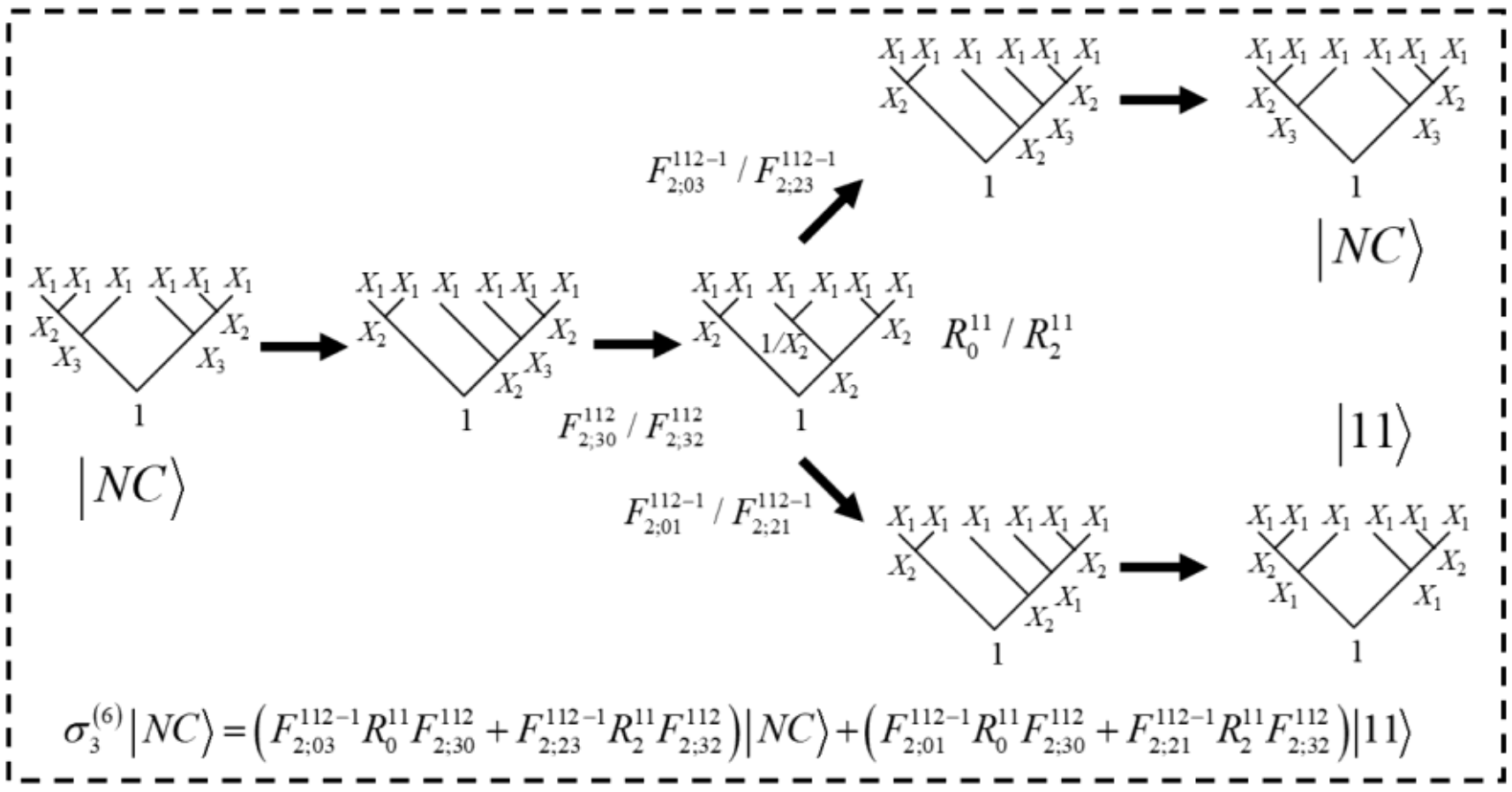

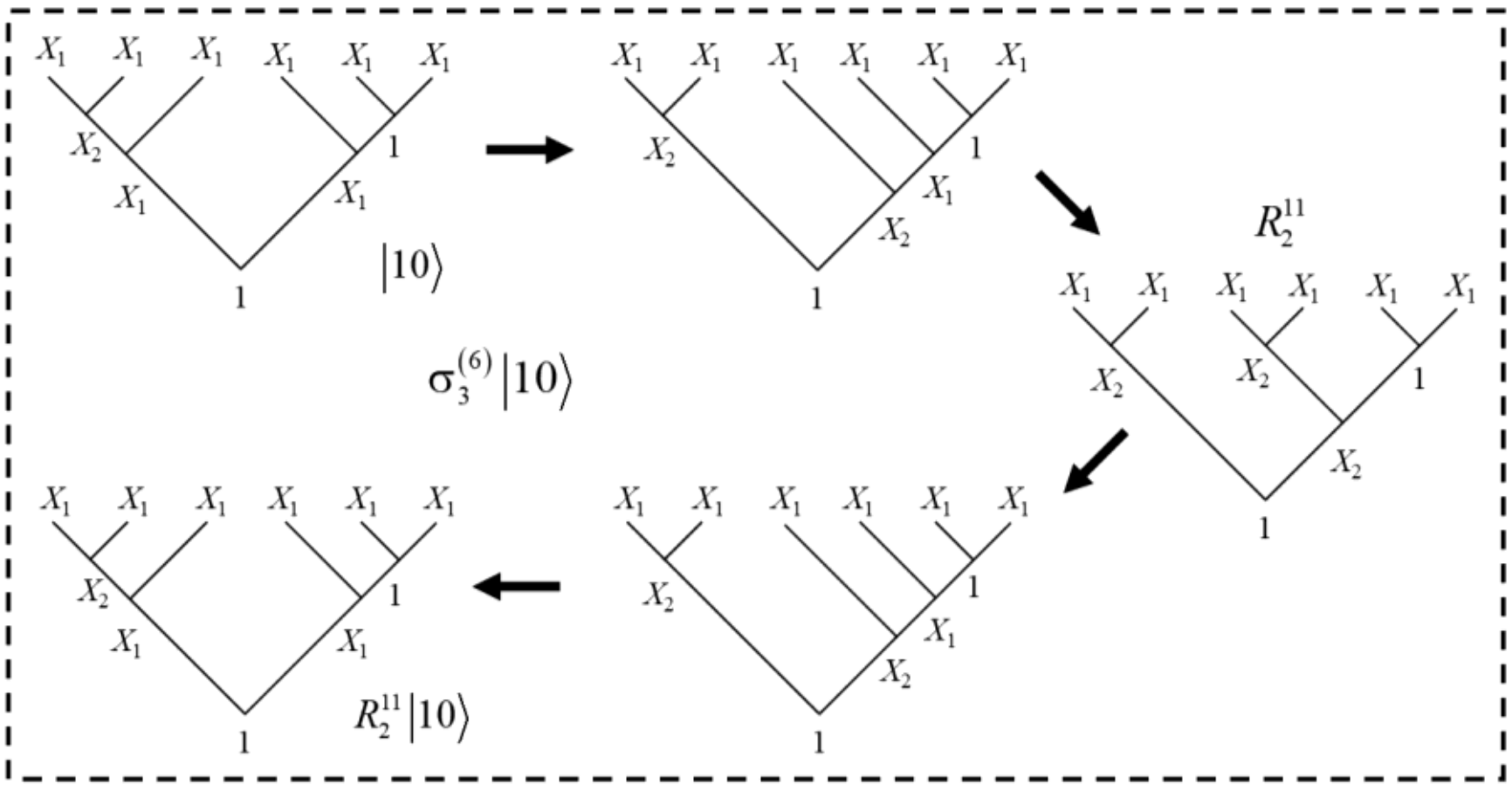

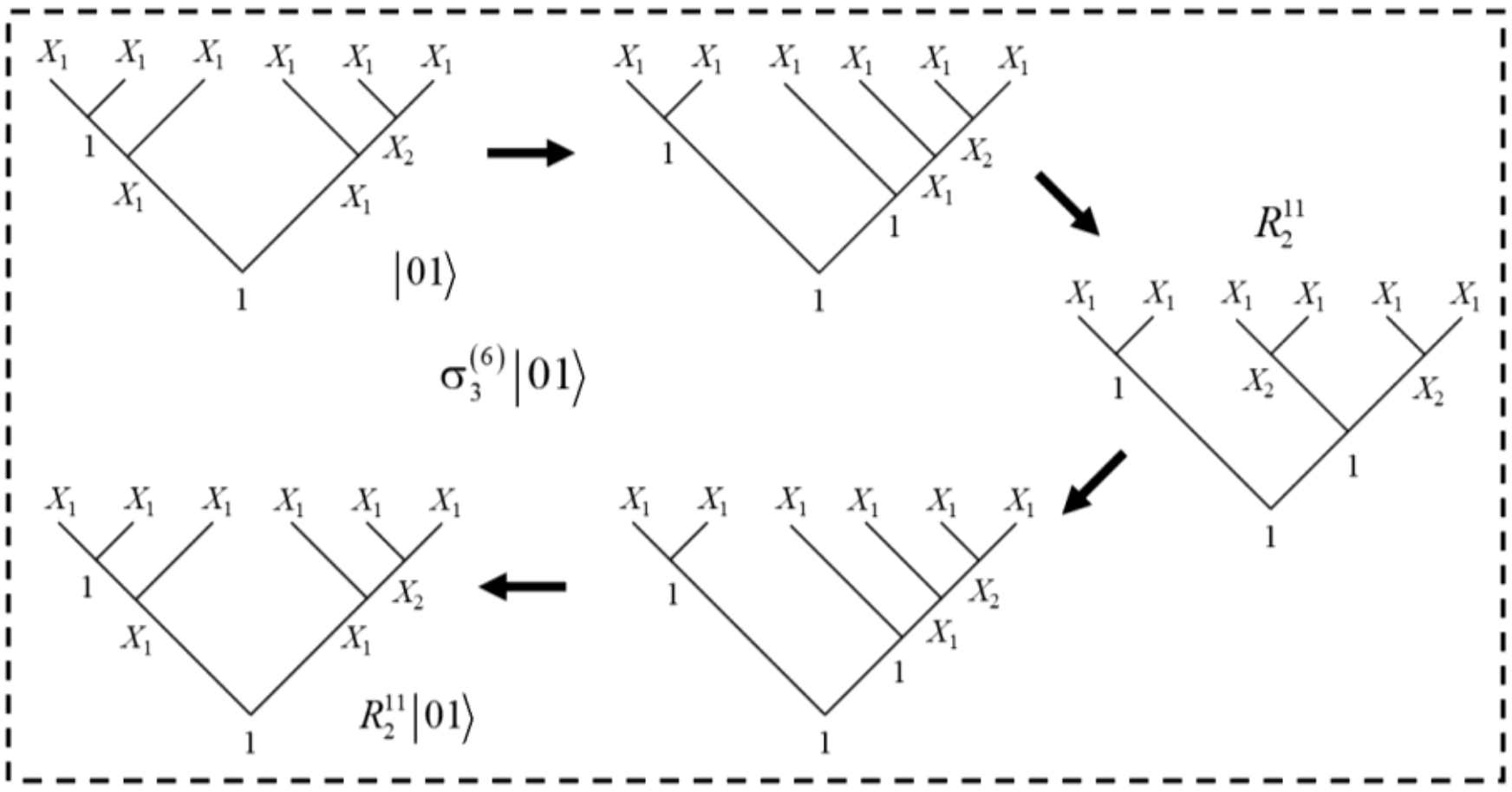

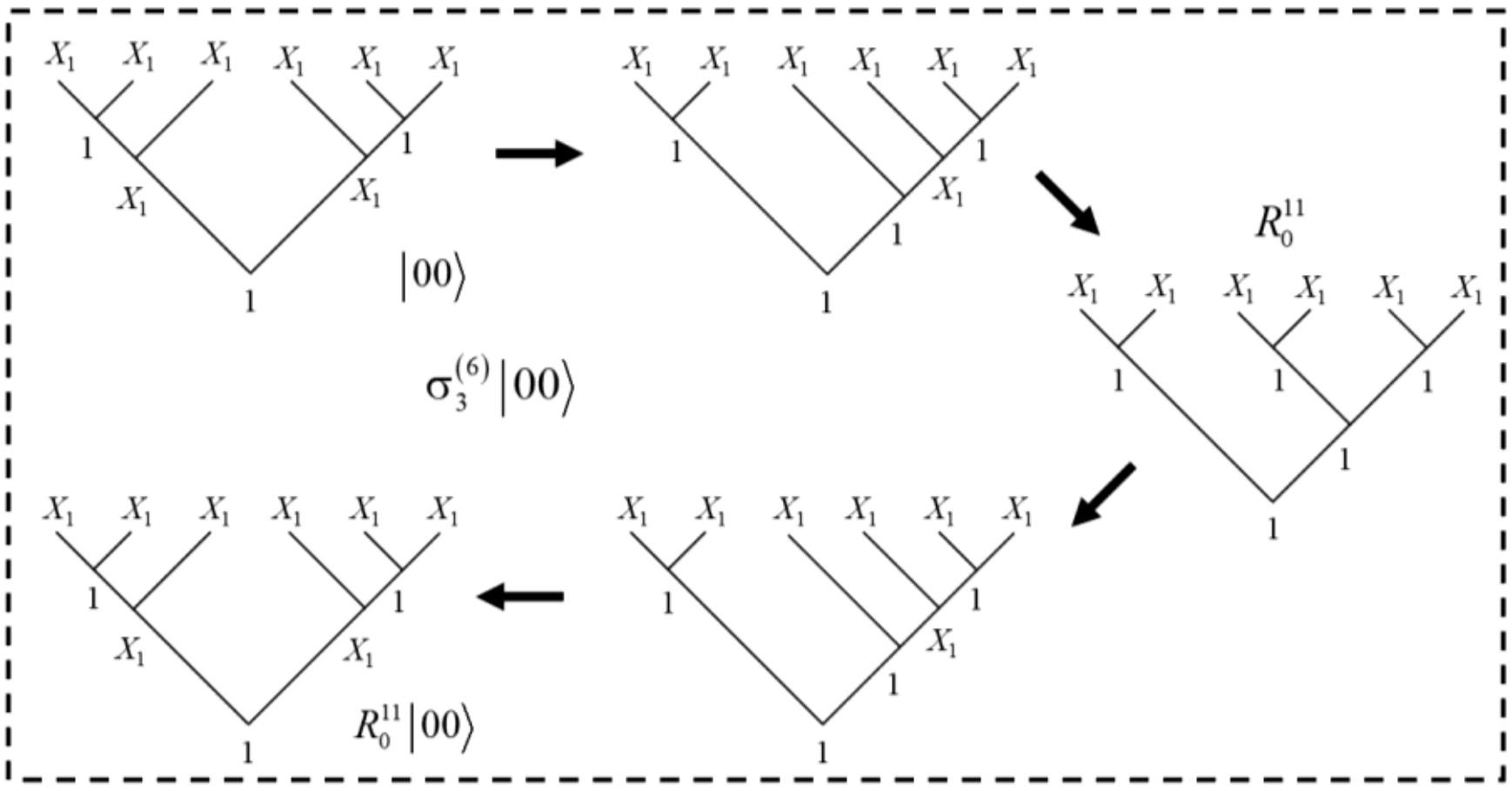

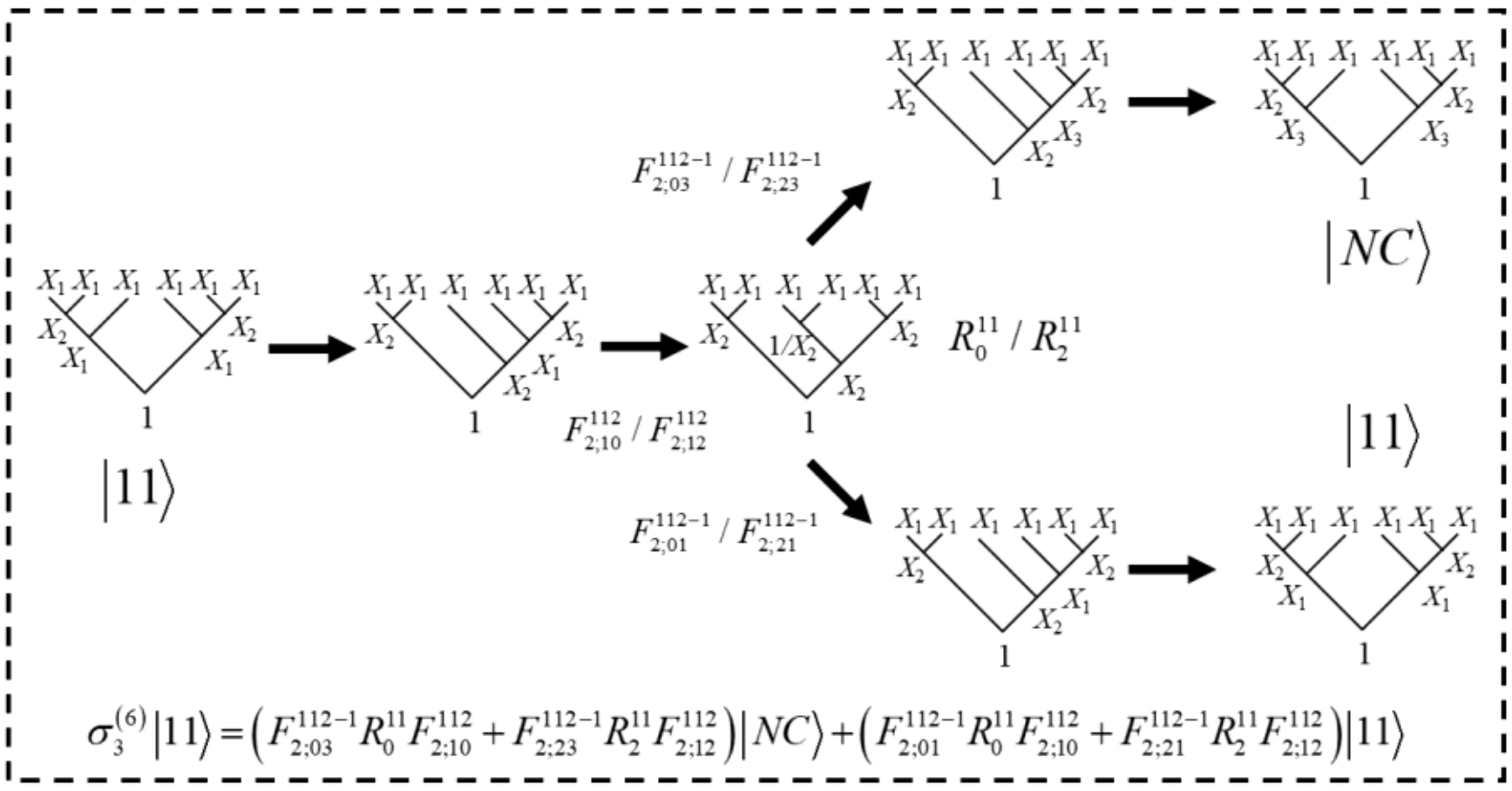

## Appendix D: Selection of topological spin particles for qubit construction

In the $SU(2)_k$ anyon model, the particle types are dictated by the level k, and their topological spins are bounded by k/2. Thus, the allowed particles with distinct topological spin labels for $SU(2)_5$, $SU(2)_6$, and $SU(2)_7$ are listed below:

$$SU(2)_5 : \frac{1}{2}, 1, \frac{3}{2}, 2, \frac{5}{2}. \quad SU(2)_6 : \frac{1}{2}, 1, \frac{3}{2}, 2, \frac{5}{2}, 3. \quad SU(2)_7 : \frac{1}{2}, 1, \frac{3}{2}, 2, \frac{5}{2}, 3, \frac{7}{2}.$$

Topological quantum computation utilizes particles with the same topological spin to form qubit. This confines our analysis to the fusion processes among identical particles. From the fusion rules Eq. (1), it follows that:

$$SU(2)_5: \frac{1}{2}\otimes\frac{1}{2}=0\oplus 1, 1\otimes 1=0\oplus 1\oplus 2, \frac{3}{2}\otimes\frac{3}{2}=0\oplus 1\oplus 2, 2\otimes 2=0\oplus 1, \frac{5}{2}\otimes\frac{5}{2}=0.$$

$$SU(2)_6: \frac{1}{2}\otimes\frac{1}{2}=0\oplus 1, 1\otimes 1=0\oplus 1\oplus 2, \frac{3}{2}\otimes\frac{3}{2}=0\oplus 1\oplus 2\oplus 3, 2\otimes 2=0\oplus 1\oplus 2,$$
$$\frac{5}{2}\otimes\frac{5}{2}=0\oplus 1, 3\otimes 3=0.$$

$$SU(2)_7: \frac{1}{2}\otimes\frac{1}{2}=0\oplus 1, 1\otimes 1=0\oplus 1\oplus 2, \frac{3}{2}\otimes\frac{3}{2}=0\oplus 1\oplus 2\oplus 3,$$
$$2\otimes 2=0\oplus 1\oplus 2\oplus 3, \frac{5}{2}\otimes\frac{5}{2}=0\oplus 1\oplus 2, 3\otimes 3=0\oplus 1, \frac{7}{2}\otimes\frac{7}{2}=0$$

Constructing qubits from anyons of the same topological spin requires the fusion outcome to be twofold. Therefore, the particles suitable for forming qubits are those with topological spins of 1/2 and 2 for $SU(2)_5$, 1/2 and 5/2 for $SU(2)_6$, 1/2 and 3 for $SU(2)_7$.

Through the calculation of EBMs, we discovered that while alternative implementations using spin-2 anyons in $SU(2)_5$, spin-5/2 anyons in $SU(2)_6$, or spin-3 anyons in $SU(2)_7$ remain viable, computational analyses confirm that the resulting EBMs differ from their spin-1/2 counterparts solely by global phase factors for $SU(2)_5$ and $SU(2)_7$, whereas the same EBMs hold for $SU(2)_6$. The results have been summarized in Table IV. Hence, anyons with a topological spin of 1/2 are used for the qubit construction in the $SU(2)_5$, $SU(2)_6$, and $SU(2)_7$ models.

**Table** IV**:** Selecting different topological spins to construct qubits yields the phase differences of EBMs based on the $SU(2)_{5,6,7}$ anyon models.

| Models | Topological spin | The phase difference of EBMs |
|---|---|---|
| $SU(2)_5$ | 1/2 or 2 | i |
| $SU(2)_6$ | 1/2 or 5/2 | same |
| $SU(2)_7$ | 1/2 or 3 | -i |

## Appendix E: The world-lines of anyons corresponding to braidword

Note that time flows from left to right, and the dot denotes an anyon with a topological spin-1/2.

The world-lines of the braidword for the *H*-gate realized by braiding operations based on $SU(2)_{3,5,6,7}$ anyon models correspond to **Table** II are shown in the figure below:

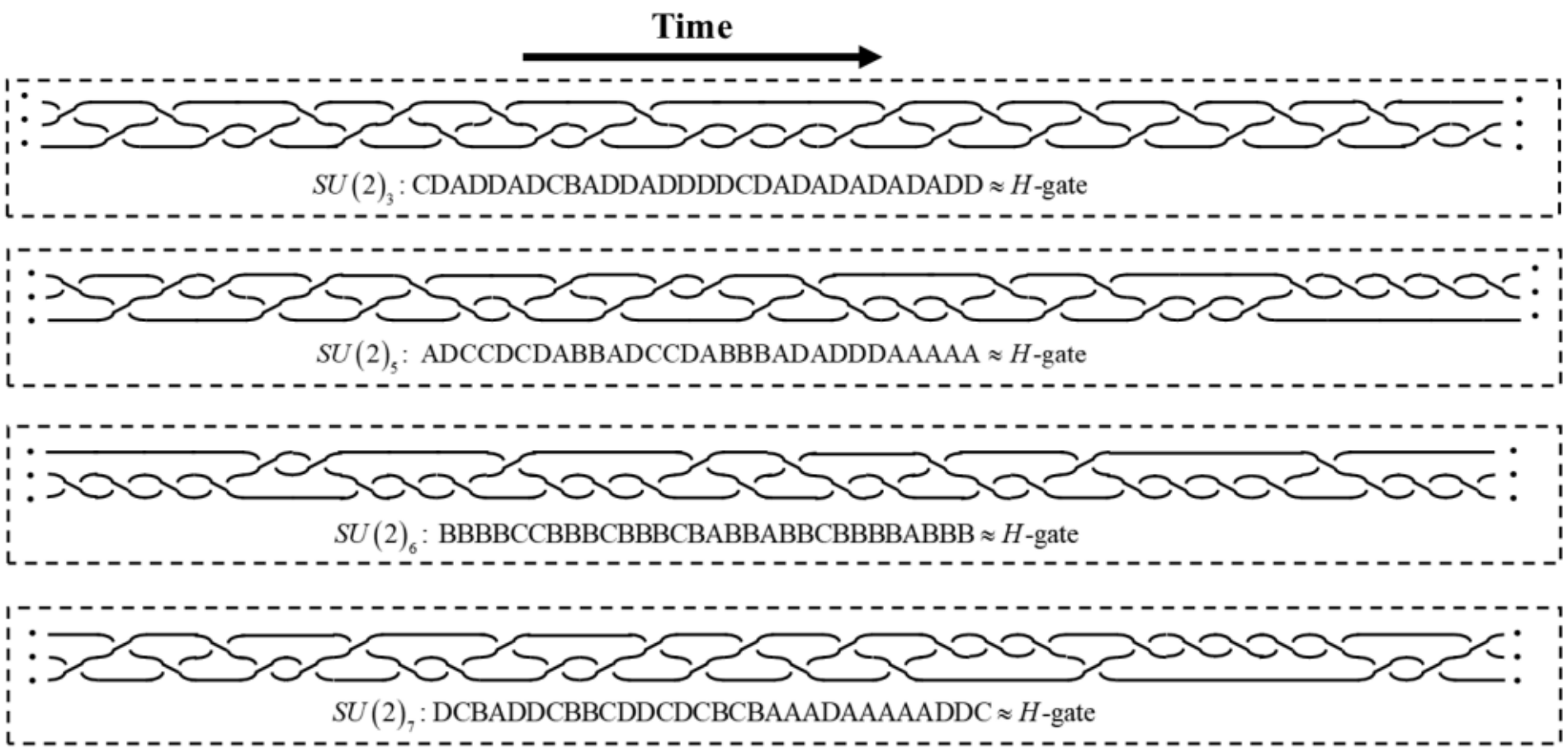

.

The world-lines of the braidword for the *T*-gate realized by braiding operations based on $SU(2)_{3,5,6,7}$ anyon models correspond to **Table** II are shown in the figure below:

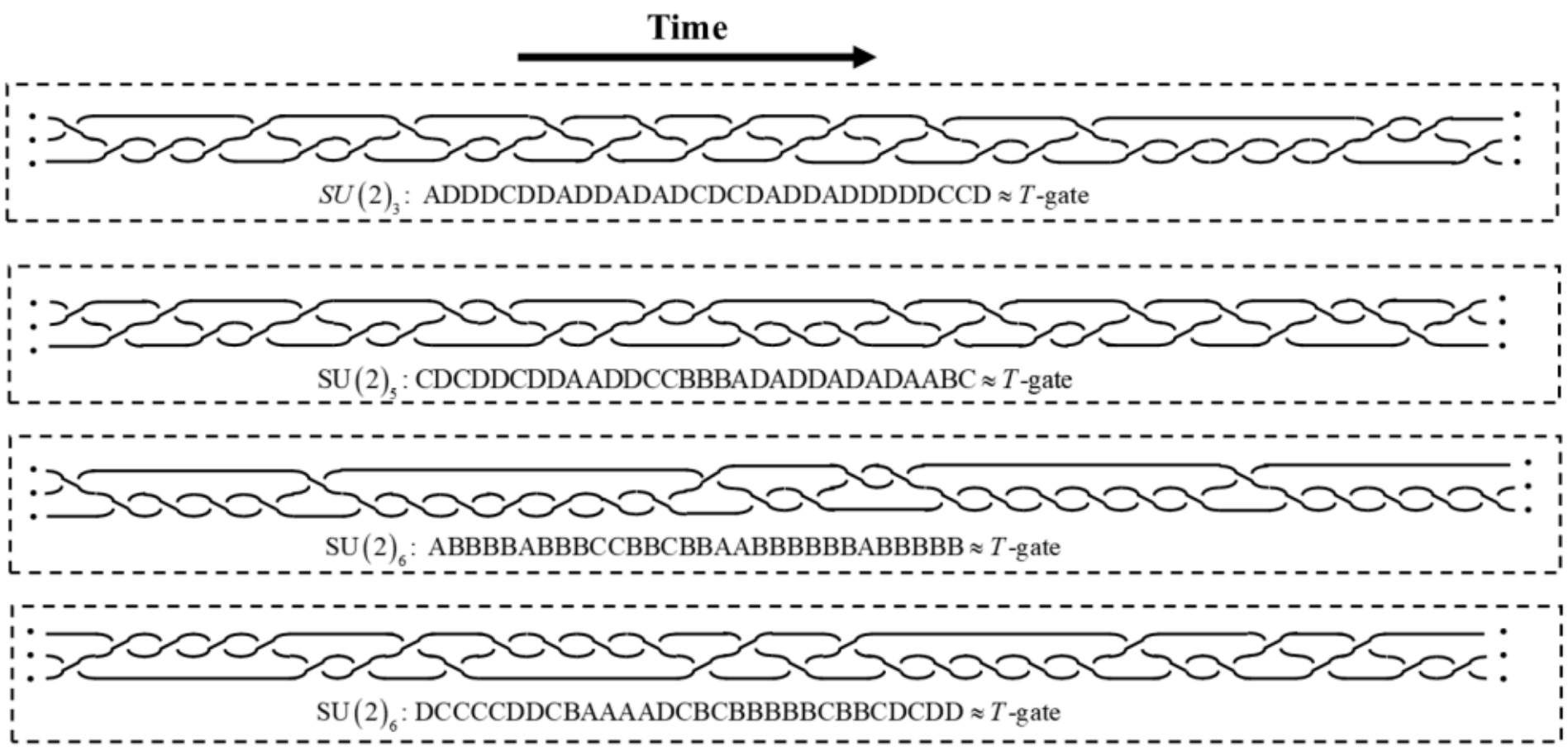

.

The world-lines of the braidword (corresponding to **Table** III, adding the inverse matrix) for the approximately local equivalence class [CNOT], which is realized by braiding operations based on $SU(2)_{3,5,6,7}$ anyon models, are shown in the figure below:

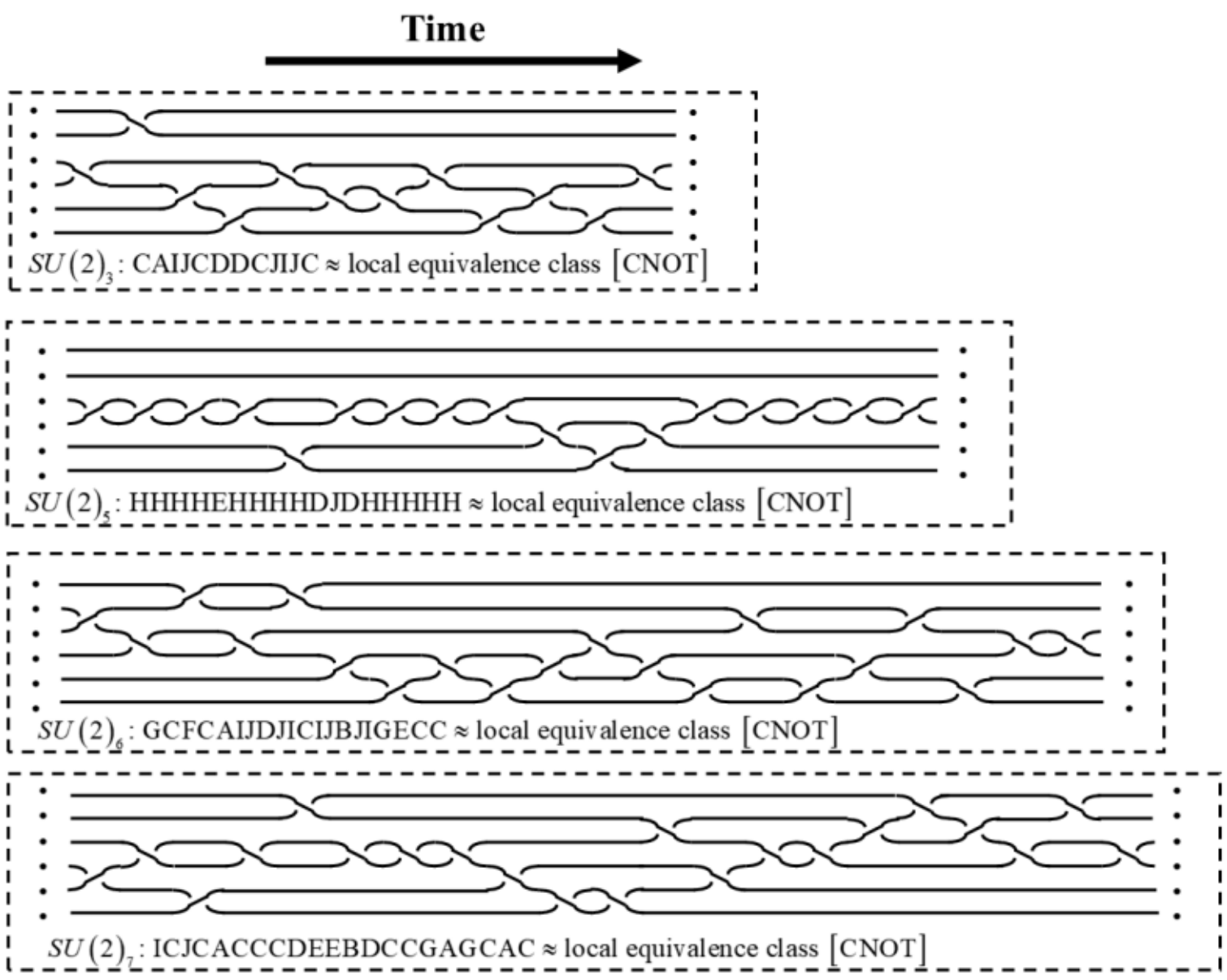

The world-lines of the braidword (corresponding to **Table** III, no inverse matrix) for the approximately local equivalence class [CNOT], which is realized by braiding operations based on $SU(2)_{3,5,6,7}$ anyon models, are shown in the figure below:

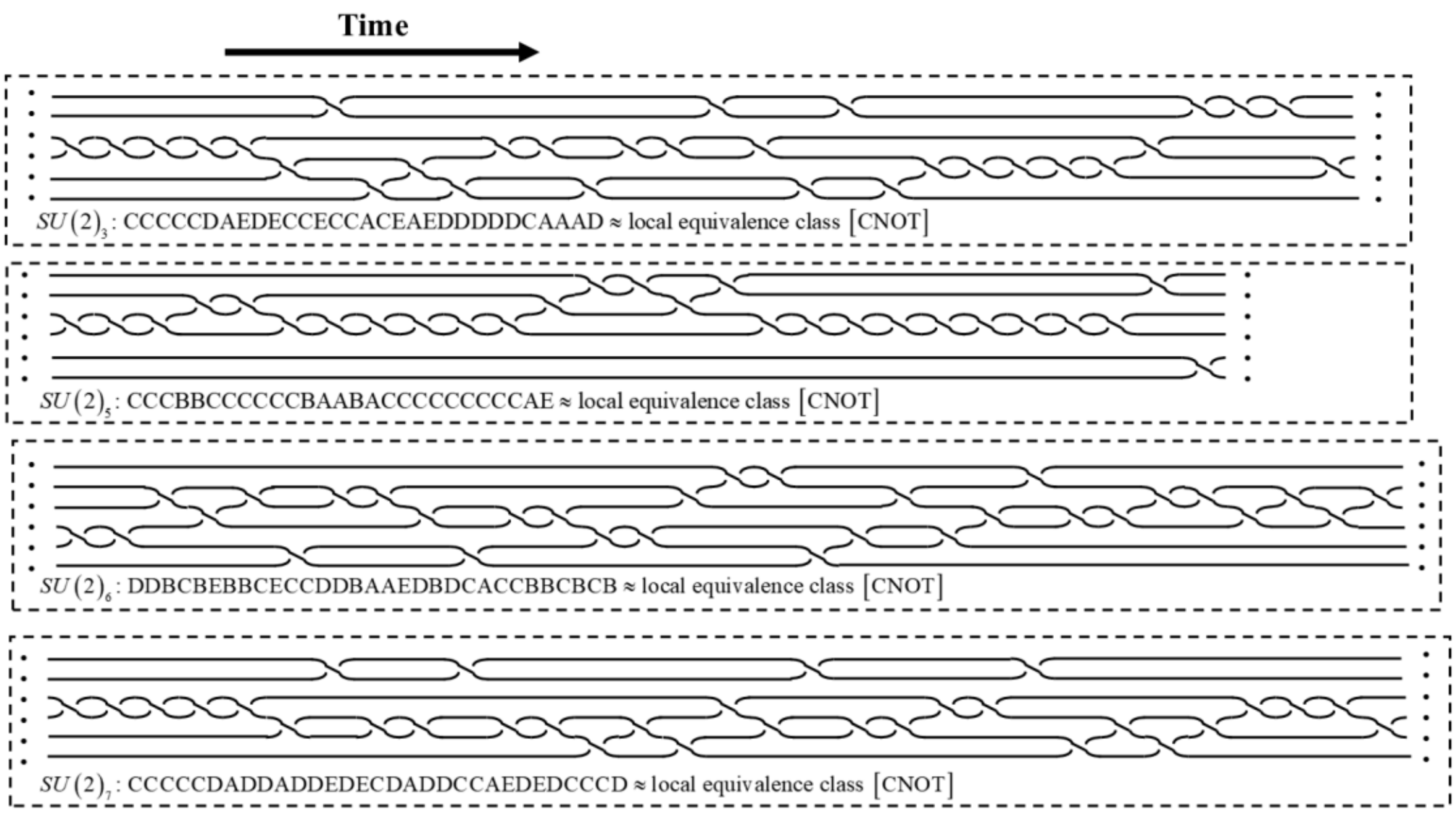

The world-lines of the braidword (corresponding to **Table** IV, no inverse matrix) for the approximately local equivalence class [SWAP], which is realized by braiding operations based on $SU(2)_{3,5,6,7}$ anyon models, are shown in the figure below:

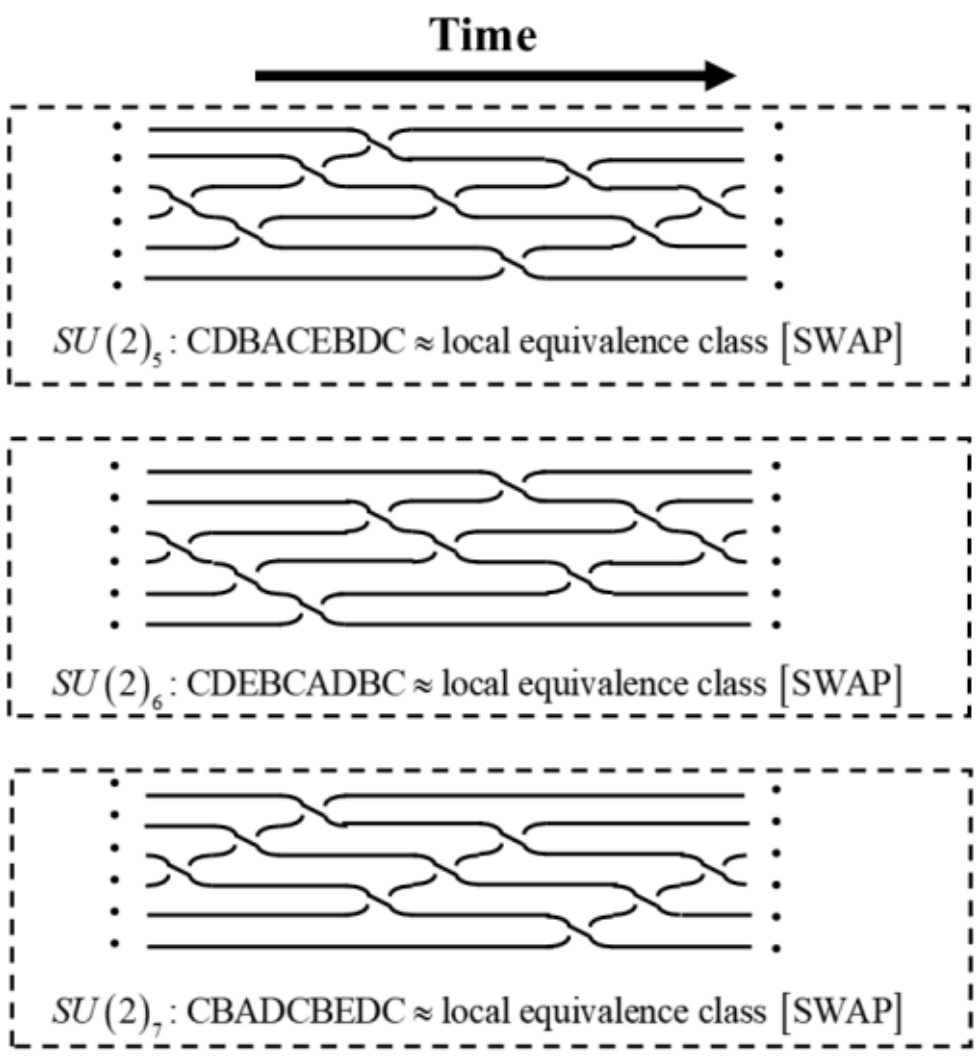